\documentclass[apj]{emulateapj}
\usepackage{natbib}
\usepackage{amsmath}
\usepackage{aas_macros}
\def\etal{{\it et al.}}

\def\MIT{\altaffilmark{1}}
\def\MITtxt{\altaffiltext{1}{MIT Kavli Institute for Astrophysics and Space Research, Cambridge, USA; clmw@mit.edu}}

\def\CfA{\altaffilmark{2}}
\def\CfAtxt{\altaffiltext{2}{Harvard-Smithsonian Center for Astrophysics, Cambridge, USA}}

\def\ASU{\altaffilmark{3}}
\def\ASUtxt{\altaffiltext{3}{Arizona State University, Tempe, USA}}

\def\ANU{\altaffilmark{4}}
\def\ANUtxt{\altaffiltext{4}{The Australian National University, Canberra, Australia}}

\def\USydney{\altaffilmark{5}}
\def\USydneytxt{\altaffiltext{5}{Sydney Institute for Astronomy, School of Physics, The University of Sydney, Sydney, Australia}}

\def\UMelbourne{\altaffilmark{6}}
\def\UMelbournetxt{\altaffiltext{6}{The University of Melbourne, Melbourne, Australia}}

\def\RRI{\altaffilmark{7}}
\def\RRItxt{\altaffiltext{7}{Raman Research Institute, Bangalore, India}}

\def\UW{\altaffilmark{8}}
\def\UWtxt{\altaffiltext{8}{University of Washington, Seattle, USA}}

\def\ICRAR{\altaffilmark{9}}
\def\ICRARtxt{\altaffiltext{9}{International Centre for Radio Astronomy Research, Curtin University, Perth, Australia }}

\def\USwinburne{\altaffilmark{10}}
\def\USwinburnetxt{\altaffiltext{10}{Swinburne University of Technology, Melbourne, Australia}}

\def\CSIRO{\altaffilmark{11}}
\def\CSIROtxt{\altaffiltext{11}{CSIRO Astronomy and Space Science, Australia}}

\def\Haystack{\altaffilmark{12}}
\def\Haystacktxt{\altaffiltext{12}{MIT Haystack Observatory, Westford, USA}}

\def\UWisc{\altaffilmark{13}}
\def\UWisctxt{\altaffiltext{13}{University of Wisconsin--Milwaukee, Milwaukee, USA}}

\def\UTasmania{\altaffilmark{14}}
\def\UTasmaniatxt{\altaffiltext{14}{University of Tasmania, Hobart, Australia}}

\def\PerthUWA{\altaffilmark{15}}
\def\PerthUWAtxt{\altaffiltext{15}{Perth Observatory, Perth, Australia, and the University of Western Australia}}

\def\CAASTRO{\altaffilmark{16}}
\def\CAASTROtxt{\altaffiltext{16}{ARC Centre of Excellence for All-sky Astrophysics (CAASTRO)}}

\begin{document}

\title{Low Frequency Imaging of Fields at High Galactic Latitude with the Murchison Widefield Array 32-Element Prototype}

%

\author{
Christopher~L.~Williams\MIT,
Jacqueline~N.~Hewitt\MIT,
Alan~M.~Levine\MIT,
Angelica~de~Oliveira-Costa\CfA,
Judd~D.~Bowman\ASU, 
Frank~H.~Briggs\ANU$^,$\CAASTRO,
B.~M.~Gaensler\USydney$^,$\CAASTRO, 
Lars~L.~Hernquist\CfA,
Daniel~A.~Mitchell\UMelbourne$^,$\CAASTRO,
Miguel~F.~Morales\UW,
Shiv~K.~Sethi\RRI,
Ravi~Subrahmanyan\RRI$^,$\CAASTRO,
Elaine~M.~Sadler\USydney$^,$\CAASTRO,
Wayne~Arcus\ICRAR,
David~G.~Barnes\USwinburne, 
Gianni~Bernardi\CfA, 
John~D.~Bunton\CSIRO, 
Roger~C.~Cappallo\Haystack,
Brian~W.~Crosse\ICRAR, 
Brian~E.~Corey\Haystack, 
Avinash~Deshpande\RRI, 
Ludi~deSouza\USydney$^,$\CSIRO,
David~Emrich\ICRAR,
Robert~F.~Goeke\MIT,
Lincoln~J.~Greenhill\CfA,
Bryna~J.~Hazelton\UW,
David~Herne\ICRAR, 
David~L.~Kaplan\UWisc, 
Justin~C.~Kasper\CfA, 
Barton~B.~Kincaid\Haystack, 
Ronald~Koenig\CSIRO, 
Eric~Kratzenberg\Haystack, 
Colin~J.~Lonsdale\Haystack, 
Mervyn~J.~Lynch\ICRAR, 
S.~Russell~McWhirter\Haystack,
Edward~H.~Morgan\MIT, 
Divya~Oberoi\Haystack, 
Stephen~M.~Ord\ICRAR, 
Joseph~Pathikulangara\CSIRO, 
Thiagaraj~Prabu\RRI, 
Ronald~A.~Remillard\MIT, 
Alan~E.~E.~Rogers\Haystack, 
Anish~A.~Roshi\RRI, 
Joseph~E.~Salah\Haystack, 
Robert~J.~Sault\UMelbourne, 
N.~Udaya~Shankar\RRI, 
K.~S.~Srivani\RRI, 
Jamie~B.~Stevens\CSIRO$^,$\UTasmania, 
Steven~J.~Tingay\ICRAR$^,$\CAASTRO, 
Randall~B.~Wayth\ICRAR$^,$\CAASTRO, 
Mark~Waterson\ANU$^,$\ICRAR,
Rachel~L.~Webster\UMelbourne$^,$\CAASTRO, 
Alan~R.~Whitney\Haystack, 
Andrew~J.~Williams\PerthUWA, 
J.~Stuart~B.~Wyithe\UMelbourne$^,$\CAASTRO
}

\shortauthors{Williams, \etal}
\shorttitle{MWA-32T Low Frequency Imaging}

\MITtxt
\CfAtxt
\ASUtxt
\ANUtxt
\USydneytxt
\UMelbournetxt
\UWtxt
\RRItxt
\ICRARtxt
\USwinburnetxt
\CSIROtxt
\Haystacktxt
\UWisctxt
\UTasmaniatxt
\PerthUWAtxt
\CAASTROtxt

\clearpage

\begin{abstract}

The Murchison Widefield Array (MWA) is a new low-frequency, wide
field-of-view radio interferometer under development at the Murchison
Radio-astronomy Observatory (MRO) in Western Australia.  We have used
a 32-element MWA prototype interferometer (MWA-32T) to observe two
$50^\circ$ diameter fields in the southern sky in the 110~MHz to
200~MHz band in order to evaluate the performance of the MWA-32T, to develop techniques for epoch of reionization
experiments, and to make measurements of astronomical foregrounds.  We
developed a calibration and imaging pipeline for the MWA-32T, and used
it to produce $\sim15'$ angular resolution maps of the two
fields.  We perform a blind source extraction using these confusion-limited
images, and detect 655 sources at high significance with an additional 871 lower significance source candidates.
We compare these sources with existing low-frequency radio surveys in
order to assess the MWA-32T system performance, wide field analysis
algorithms, and catalog quality.  Our source catalog is found to agree
well with existing low-frequency surveys in these regions of the sky and with statistical distributions of point
sources derived from Northern Hemisphere surveys; it
represents one of the deepest surveys to date of this sky field in the
110~MHz to 200~MHz band.

\end{abstract}


\section{Introduction}

The study of the origin and evolution of the Universe draws upon
observations of phenomena at a large range of distances and look-back
times, connecting the initial conditions probed by the cosmic
microwave background to present-day conditions dominated by galaxies
and clusters of galaxies.  A major chapter of this history has yet to
be examined --- the chapter that corresponds to redshifts from
$z=1000$ to $z=6$ and comprises the Dark Ages and the Epoch of
Reionization (EoR) of the Universe.  The EoR, in particular, marks a major milestone
when the first stars and galaxies formed and reionized the
intergalactic medium.  It was recognized some time ago that studies of
the redshifted 21-cm radio emission from neutral hydrogen would be a
promising probe of the EoR \citep{Hogan1979,Scott1990,Madau1997}.
Indeed, the possible existence of extensive regions containing significantly large amounts of neutral hydrogen at redshifts of $z=15$ to $z=8$
 motivates an interest in developing highly sensitive
low-frequency radio telescopes in order to detect the redshifted 21-cm
signal.  Not long ago it was recognized that a statistical detection
of the patchy neutral hydrogen distribution during the EoR should be
possible with an array with a modest collecting area and a large field
of view \citep{Morales2004}.  In the past decade considerable efforts have been made towards this goal, through advances in theoretical modeling of the EoR signature (see e.g. \citealt{Pritchard2011}, \citealt{Morales2010},  and \citealt{Furlanetto2006} for recent reviews), as well as through the development of new instrumental approaches to measure the redshifted 21-cm signal (see e.g. \citealt{Bowman2010}, \citealt{Chippendale2009}, \citealt{Lonsdale2009}, \citealt{Tingay2012}, \citealt{Parsons2010}, \citealt{Rottgering2006}, and \citealt{Paciga2011}).

One particularly daunting challenge for these experiments is emission from foreground astrophysical sources, which is at least two to three orders of magnitude brighter than the redshifted 21-cm signal (in the total intensity as well as in the magnitude of the spatial fluctuations, see e.g. \citealt{Shaver1999, Bernardi2009, Pen2009, DeOliveiraCosta2008}).  The foreground emission arises mainly from
synchrotron and free-free processes, and therefore is highly likely to
have an intrinsically smooth radio spectrum.  The 21-cm signal,
however, is likely to be produced under conditions that vary rapidly
with redshift, and, if this is the case, will appear to have rather
sharp spectral features.  Techniques have been developed to
exploit this differing spectral behavior in order to separate and
remove the foreground contamination \citep{Furlanetto2004,McQuinn2006,Bowman2009,Harker2009,Liu2009,Liu2011}.  However,
any calibration imperfection or instrumental defect has the potential
to introduce distortions into measured spectra, and thereby to mix the
extremely bright foreground emission with the signal from the
redshifted 21-cm signal in ways that are difficult to disentangle \citep{Datta2010}.  Thus, an
equally daunting challenge is to learn how to calibrate any new
instrumentation that is being developed for these observations with
extremely high fidelity.

The Murchison Widefield Array  (MWA,
\citealt{Lonsdale2009, Tingay2012}) is a new array being constructed to characterize the 21-cm signal during the EoR\footnote{Additionally, the MWA has been designated an official precursor instrument for the Square Kilometre Array (SKA).}.  In addition to the study of the EoR, other
key science goals of the MWA include the study of radio transients, the study of the heliosphere and ionosphere, and low-frequency Galactic and extragalactic studies.  These four goals
and potentially others are addressed by an array made of a large
number of small antenna elements that simultaneously give a large
collecting area and a large field of view.  This is a departure in many ways from a traditional radio array
design, with phased arrays of dipoles constituting the
fundamental antenna elements, digitization early in the data stream,
and full correlation of a large number of baselines.  This design promises large improvements with regard to
wide-field surveys and detection of the EoR, but, at the same time, it
poses new challenges, especially with regard to calibration and imaging of the
large field of view and compensation for the effects of the
ionosphere. The instrument is currently under construction; work on a
128-element array commenced in early 2012 at the radio-quiet Murchison
Radio-astronomy Observatory (MRO) in Western Australia.  As a first
step in demonstrating the new technologies required for the MWA, a
32-element prototype was built at the MRO site prior to the build-out
to 128 elements.  This prototype system was operated for two years in
campaign mode, and underwent a cycle of equipment installation,
testing, and redesign as necessary.  Beginning in March 2010, the
prototype was used for initial science observations, and has already
yielded several results.  \citet{Oberoi2011} presents findings
from an investigation of solar radio emission, and \citet{Ord2010}
present wide-field images using a prototype real-time imaging and calibration pipeline.  Herein, we
also report results based on data obtained during  this initial science run.

The goals of the measurements and analysis presented here are to
verify the performance of the MWA subsystems and the 32-element
prototype array, to explore techniques for future EoR experiments, and
to deepen our understanding of the astronomical foregrounds.  We
observed two overlapping fields at high Galactic latitude, each
$50^\circ$ across.  One field was identified for possible EoR studies
in the future, and the other was chosen to have a very bright radio
source at its center to facilitate calibration.  The observations are
deep in the sense that they combine data from a large range of hour angles and multiple snapshot images
to improve sensitivity and image fidelity; developing such
techniques for deep observations is critical for the success of future
EoR experiments and other scientific investigations with the MWA.  We
have developed a data reduction pipeline that implements wide-field
calibration and imaging algorithms, compensates for the
direction-dependent and changing primary beam as different snapshots
are combined, and automatically extracts sources from the images.  We
compare our results directly to the results from other sensitive
low-frequency radio surveys in the southern hemisphere, and we compare
them statistically to the results from surveys carried out in the
northern hemisphere.  We use these comparisons to assess the
performance of the MWA prototype and the wide-field imaging and
calibration algorithms.  We assess the completeness and reliability of
our point source catalog through comparison to these surveys.  We make
a number of simplifying assumptions in this first phase of analysis;
future work will refine the techniques and algorithms until the
stringent calibration requirements of EoR experiments with the full
MWA can be met.

\section{Low-Frequency Radio Surveys}
\label{sec:surveys}

In this work we make extensive use of the results of previous
sensitive low-frequency radio surveys to verify the performance of the
MWA and to provide external data for calibration.  We summarize here
the properties of the surveys used in our comparisons.

The Molonglo Reference Catalog (MRC; \citealt{Large1981}) is the
product of a blind survey at 408 MHz that covered nearly all of the
southern sky to moderate depth. The catalog covers all right
ascensions in the declination range from $-85.0^\circ$ to
$+18.5^\circ$, excluding the area within $3^\circ$ of the Galactic
plane.  The observations were conducted with the Molonglo Radio
Telescope in a 2.5~MHz wide band with a synthesized beam
of $2.62' \times 2.86' \sec(\delta+35.5^\circ)$ in width.  The MRC has a stated
completeness limit of 1~Jy, although it contains sources down to a
flux of $\sim$0.7~Jy.

At frequencies below 408~MHz, there have been many targeted
observations of known sources in the southern sky.  The Culgoora
Circular Array \citep{Slee1995} was used to observe Galactic and
extragalactic sources selected from existing higher frequency surveys.
The observations were made at frequencies of 80~MHz and 160~MHz.  The
beam size was $3.70' \times 3.70' \sec(\delta+
30.3^\circ)$ at 80~MHz, and $1.85' \times 1.85'
\sec(\delta+30.3^\circ)$ at 160~MHz.  The limiting flux density was
4~Jy at 80 MHz, and 2~Jy at 160~MHz. However, only a small patch of
sky around each selected source was imaged.  Although flux density uncertainties are not directly reported in this list, \citet{Slee1977} note that the standard deviation in the flux density for sources measured with the Culgoora array is $\sim13\%$ for the brightest sources, and $\sim39\%$ for the faintest sources, with a potential systematic flux scale depression of $\sim10\%$.  Similarly,
\citet{Jacobs2011} present results from PAPER, an array of east-west
polarized dipoles, that were obtained over the 110~MHz to 180~MHz
band.  The results were derived from multi-frequency synthesis maps of
the entire sky south of a declination of $10^\circ$ having a
resolution of 26$'$.  A sample of 480 sources with fluxes
greater than 4~Jy in the MRC were identifed and measured in these
PAPER maps. \citet{Jacobs2011} find a 50\% standard deviation in their fluxes relative to values obtained from the MRC and Culgoora source lists. They quote
a flux limit of 10~Jy for the sources in their catalog.

There are several ongoing efforts to perform low-frequency blind
surveys.  \citet{Pandey2006} presents results from a survey that used
the Mauritius Radio Telescope (MRT) to image $\sim$1 steradian of the
sky at 151~MHz and to thereby produce a catalog of 2782
sources\footnote{Electronic catalogs are available at {\tt
http://www.rri.res.in/surveys/MRT}}.  The deconvolved images achieve
an angular resolution of $4' \times 4.6' \sec(\delta +
20\fdg14)$ and a root-mean-square (RMS) noise level of approximately
$300~{\rm mJy}~{\rm beam}^{-1}$ \citep{Nayak2010}.  The TIFR GMRT Sky
Survey (TGSS, \citealt{Sirothia2011}) is producing a 150~MHz survey of
the sky at declinations above $-30^\circ$.  Each pointing covers
$\sim7$ square degrees and yields a map that reaches an RMS noise of
$\sim8~{\rm mJy}~{\rm beam}^{-1}$ at an angular resolution of $\sim20''$.  The
flux density scales of the maps are limited by systematic errors and
have relative errors of 25\%.  As of 2012
January, the TGSS website\footnote{{\tt http://tgss.ncra.tifr.res.in/150MHz/tgss.html}} reports results from images of approximately
2600 square degrees of the southern sky.

Surveys which primarily cover the northern sky have also been carried out at low
frequencies.  The most extensive wide-field uniform survey near our
observing frequency of 150 MHz is the 6C survey
\citep{Baldwin1985,Hales1988,Hales1990,Hales1991,Hales1993a,Hales1993b}.
The 6C survey covered the northern sky above declination $30^\circ$
with a sensitivity of 200 mJy; the angular resolution was $4.2' \times
4.2' \csc \delta$.  The 7C survey \citep{Hales2007} covers 1.7~sr of
the northern sky to a greater depth and at higher resolution than the
6C survey.  We have chosen to use the 6C survey for comparison to our
results in this paper because it covers a somewhat larger sky area, and has served as the basis for other investigations of EoR foregrounds (in particular, \citealt{DiMatteo2002}).  At lower frequencies, \citet{Cohen2007} have used the VLA 74~MHz system to perform a survey of the sky north of declination $\delta=-30^\circ$.  This survey, known as the VLA Low-Frequency Sky Survey (VLSS), produces maps with an $80''$ angular resolution which achieve a typical RMS noise level of $100~\rm{mJy}~\rm{beam}^{-1}$.  Each VLSS image is $14^\circ \times 14^\circ$ across in order to fully image the VLA primary beam, which has a FWHM diameter of $11.9^\circ$ degrees.  \citet{Cohen2007} perform a blind source extraction on the VLSS maps and produce a source catalog of 68,311 radio sources above a significance level of 5$\sigma$.  They quote a 50\% point source detection limit of $0.7~{\rm Jy}~{\rm beam}^{-1}$.

For completeness we note that additional low-frequency surveys include the
Sydney University Molonglo Sky Survey (843~MHz, \citealt{Bock1999}),
the Miyun survey (232~MHz, \citealt{Zhang1997}), and the Levedev
Physical Institute Survey (102.5~MHz, \citealt{Dagkesamanskii2000}).
We have not extensively compared our results to results from these surveys.

Low frequency surveys have also been used to study the distribution of
spectral indices of radio sources. \citet{DeBreuck2000} used
results from the MRC and from the Parkes-MIT-NRAO 4.85 GHz survey
(PMN,
\citealt{Wright1994,Griffith1993,Griffith1994,Condon1993,Tasker1994,Griffith1995,Wright1996})
to study the distribution of spectral indices of sources in the
southern sky.  They also carried out similar comparisons of the
results from the Westerbork Northern Sky Survey (325 MHz;
\citealt{Rengelink1997}) and the Texas Survey (365 MHz;
\citealt{Douglas1996}) with results from the NRAO VLA Sky Survey (1.4
GHz; \citealt{Condon1998}) in the northern hemisphere.  The spectral
index distributions showed significant differences between samples
selected at low frequencies and samples selected at high frequencies.
The combined MRC-PMN source list was also used to generate a
sample of ultra-steep spectrum sources.

In Section~\ref{sec:results}, we carry out source-by-source comparisons of our
survey results to those of the MRC, the Culgoora flux density
measurements, the PAPER flux density measurements, and the TGSS.
There is no overlap at present between our survey and the Mauritius
survey, but  comparisons should become possible when the analysis of the Mauritius 
data is completed.  We also carry out statistical
comparisons of our survey results to those of the 6C survey by
comparing source counts, and to the \citet{DeBreuck2000} spectral
index catalogs by comparing spectral index distributions.

\section{The MWA-32T Instrument}

The Murchison Widefield Array 32-Tile prototype (MWA-32T) was built
and operated for the purpose of verifying the performance of MWA
subsystems in preparation for building a larger, more capable array.
As noted above, construction of a 128-tile array has commenced and is
expected to be complete later this year (2012).  We summarize the
design here; the reader is referred to \citet{Lonsdale2009} and
\citet{Tingay2012} for more detailed descriptions.

The MWA-32T was designed to cover a frequency range from 80~MHz to
300~MHz, with an instantaneous bandwidth of 30.72~MHz at a spectral
resolution of 40~kHz.  The array consisted of 32 antenna tiles, which
served as the primary collecting elements of the array.  The tiles are
designed to have an effective collecting area larger than $10~{\rm
m}^2$ in the MWA frequency band, and to provide a steerable beam which
can be pointed up to $60^\circ$ from zenith while maintaining a
system temperature which is dominated by sky noise within the MWA
band.  The primary beam of the tiles is frequency and position
dependent, with a FWHM size at zenith of roughly $25^\circ /
(\nu/150~{\rm MHz})$.  Each tile consists of 16 dual-polarization,
active dipole antennas laid out over a metal mesh ground screen in a $4\times4$ grid with a 1.1~meter
center-to-center spacing.  Each dipole antenna consists of vertical
bowtie elements that feed a pair of integrated low-noise amplifiers
(LNAs) located within a tube at the juncture of the
orthogonal arms of the dipole.  The antennas are designed to have low
horizon gain to reduce terrestrial RFI contamination, and to have a low
manufacturing cost.

The signals for the two polarizations are processed in parallel.  For
each polarization, the signals from the 16 dipoles on each tile are carried over
coaxial cable to an analog beamformer, where they are coherently
summed to form a single tile beam.  A system of switchable analog
delay lines is used to apply an independent time delay to each of the
dipole signals, allowing the tile beam to be steered on the sky.  The
delay lines employ a series of 5 switchable traces, each differing by
a factor of two in length, with the shortest trace introducing a
nominal delay of 435~ps.  This allows for 32 discrete delay settings
for each of the input signals.  The discretization of the delays
implies that the primary beam can only be steered in discrete steps,
and so can only coarsely track a sky field.  The summed signal is
amplified and sent over coaxial cables to the MWA digital receiver for
digitization.

Each MWA digital receiver node services 8 tiles.  The 16 received
signals are first subjected to additional filtering and signal
conditioning for low-frequency rejection, anti-aliasing, and
level adjustment.  The signals are then digitized at baseband by eight
dual 8-bit analog-to-digital converter (ADC) chips operating
at a sampling rate of 655.36~MHz.  The data stream from each ADC is fed to a digital
polyphase filterbank (PFB) implemented in FPGA hardware which
produces 256 frequency channels, each 1.28~MHz wide.  A sub-selection
of 24 of these channels (a bandwidth of 30.72~MHz) are transmitted via
optical fiber to the correlator.

At the correlator, the data streams from each receiver are processed by a second stage PFB to
obtain a frequency resolution of 10~kHz.  The signals are then cross-multiplied to produce a 3072 channel complex spectrum
for each of the 2080 correlation products. These comprise the 
four polarization products for all pairs of tiles as well as the
autocorrelations.  The visibilities are
averaged into 40~kHz wide channels and integrated for 50~ms due to
output data rate constraints.  During
the 32T observing campaign described in this work, the correlator was
operating at a 50\% time duty cycle due to hardware limitations.  The
visibilities are captured and averaged into 1~second integrations
before being written to disk.

\section{Observations}

Observations were conducted with the MWA-32T in March 2010 during a
two-week campaign (X13) when personnel were present on-site to operate the
instrument.  Data were taken in three 30.72~MHz sub-bands centered at
123.52~MHz, 154.24~MHz, and 184.96~MHz in order to give (nearly)
continuous frequency coverage between $\sim$110~MHz and $\sim$200~MHz.
During the observations, the beamformers were used to steer the beam
in steps as the fields crossed the sky.  This stepped steering is a
consequence of the discretization of the analog delay lines in the
beamformer.  The typical sequence was to steer the beam to a new
position, observe at a particular frequency for five minutes (without
tracking), and then steer the beam again. Thus, the measurements can
be considered to be a series of short drift scans.

The observing time was divided between two fields.  One field was
centered on the bright extragalactic source Hydra~A at RA(J2000) $ =
9^{\rm h}\ 18^{\rm m}\ 6^{\rm s}$, Dec.(J2000) $= -12^\circ\ 5'\
45''$ to facilitate calibration.  The other covered the
EoR2 field, centered at RA(J2000) $= 10^{\rm h}\ 20^{\rm m}\ 0^{\rm s}$,
Dec.(J2000) $= -10^\circ\ 0'\ 0''$.  The EoR2 field is one
of two fields at high Galactic latitude that have been identified by the MWA
collaboration as targets for future EoR experiments.  It also had the
advantage of being above the horizon at night during the observing
campaign.  Although the centers of the Hydra~A and EoR2 fields are
separated by $15.3^\circ$, there is considerable overlap between them
since the half power beam width of the primary beam is $\sim25^\circ$
at 150~MHz.  A total of 61 $\sim$5 minute scans of the Hydra~A field and
248 scans of the EoR2 field were obtained in interleaved sequences
over the course of the observing sessions. Table~\ref{tab:obsjournal}
gives a journal of the observations.

\begin{deluxetable}{lcccc}
\tablecaption{Journal of Observations}
\tablehead{ \colhead{Field} &  \colhead{Frequency}    &   \colhead{Date}  &   \colhead{Number} & \colhead{Observation\tablenotemark{a}}  \\
& \colhead{(MHz)} & &  \colhead{of Scans} & \colhead{Time (minutes)} \\
}
\tablewidth{0pt}
\startdata
EoR2      & 123.52   & 2010/03/24  & 35 & 208 \\
                &         & 2010/03/28  & 26 & 154 \\
                & 154.24 & 2010/03/22  & 35 & 208 \\
                &         & 2010/03/26  & 18 & 107 \\
                &         &  2010/03/28 & 10 &  59  \\
                &  184.96 &  2010/03/21 & 35 & 208 \\
                &         & 2010/03/25  & 35 & 208 \\
                &         & 2010/03/26  & 18 & 106 \\
                &         & 2010/03/29  & 36 & 214 \\	
Hydra~A & 123.52    & 2010/03/24  & 8 &  39 \\
                &         & 2010/03/28  & 7 &  34 \\
                & 154.24 & 2010/03/22  & 8 & 39\\
                &         & 2010/03/26  & 5 &  24\\
                &         &  2010/03/28 & 3 & 15\\
                & 184.96 &  2010/03/21 & 8 & 39\\
                &         & 2010/03/25  & 8 & 39\\
                &         & 2010/03/26  & 5 & 24\\
                &         & 2010/03/29  & 9 & 44\\
\enddata
\tablenotetext{a}{The effective integration time is less than half of this observation time due to the 50\% duty cycle of the correlator and additional flagging.}
\label{tab:obsjournal}
\end{deluxetable}

\section{Data Reduction Strategy\label{sec:Reduction}}

\subsection{Instrumental Gain Calibration \label{sec:CalStrategy}}

The MWA antenna tile architecture poses several nontraditional calibration issues due to
both the nature of the primary beam and the wide field-of-view.  The
primary beam is formed by the summation of beamformer-delayed
zenith-centered dipole responses.  The beamformer delays are
periodically changed to track a field across the sky.  Although this moves the
center of the primary beam as intended, the overall shape of the beam changes
as well.  For a given set of beamformer delays, the beam is fixed
relative to the tile, and therefore moves relative to the sky as the
Earth rotates. As a further complication, the time and direction
dependence of the primary beam response is different for the two
polarizations of the crossed dipoles in the array.  This leads to
apparent  polarization in inherently unpolarized
sources due to the different responses in the two
orthogonal dipole polarizations unless the appropriate corrections are applied in the analysis procedure.

Methods for measuring and calibrating the primary beam have been developed for the full MWA system by \citet{Mitchell2008}, who plan to use a real-time system (RTS) to calibrate and image MWA data.  Their method performs a calibration of the instrument in real time, solving for direction- and frequency-dependent gains for each antenna based on the simultaneous measurement of multiple known bright sources across the field of view.  This method was developed for use in a 512-tile array, where the instantaneous sensitivity and $uv$-plane coverage enable the measurement of several hundred sources in each 8-second iteration \citep{Mitchell2008}.  \citet{Ord2010} have successfully demonstrated  the use of a modified version of the RTS in order to calibrate and image data from the MWA-32T.  However, the reduced sensitivity of the MWA-32T array makes this full calibration challenging.  For the MWA-32T system, the data rate is sufficiently low that real-time calibration and imaging are not necessary, and the raw visibility data can be captured and stored.  We have chosen to pursue an alternative data reduction pipeline based on more traditional calibration and imaging software which allows us to use the full visibility data set in order to perform a detailed investigation of the calibration and imaging performance of the MWA-32T instrument.

Without the ability to directly measure the primary beam for each
tile, we instead assume a model and use it to account for the
instrumental-gain direction dependences. Knowledge of the primary
beams is also needed for an optimized weighting of the maps when
combining them to obtain deeper maps (see
Section~\ref{sec:weighting}).  It is likely that precise
characterization of individual tile beams will be necessary to achieve
dynamic range sufficient for accurate foreground subtraction and EoR
detection, but this is not attempted in the work described herein.

\begin{figure}
\plottwo{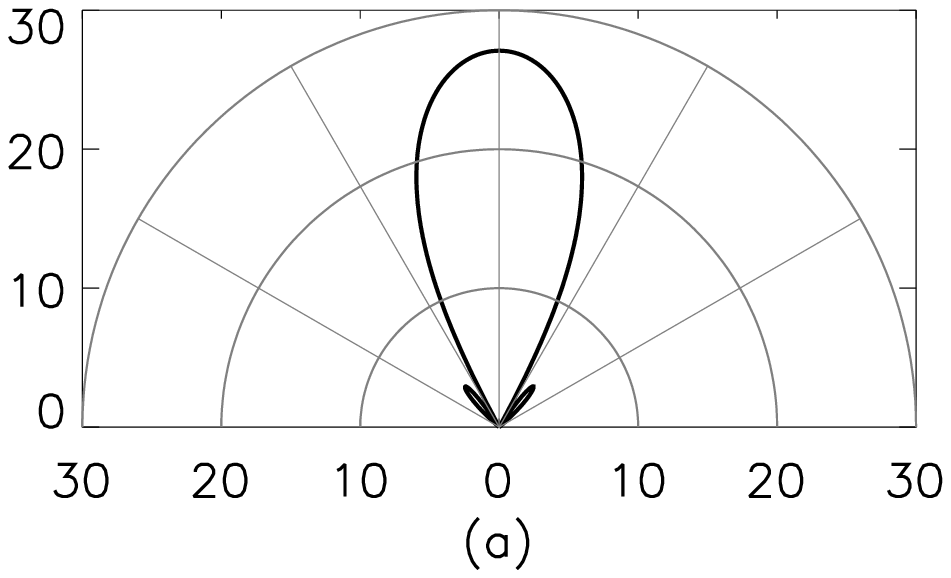}{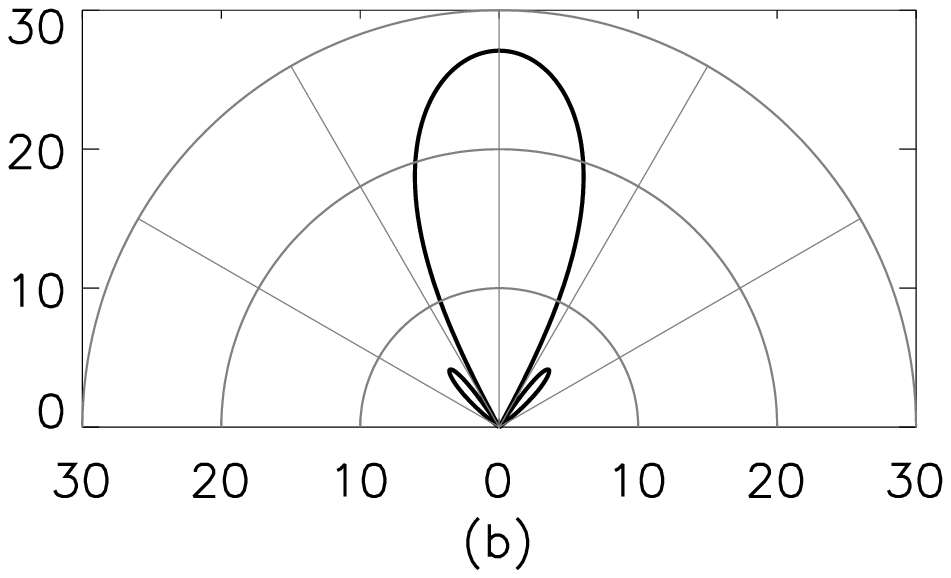}\\
\plottwo{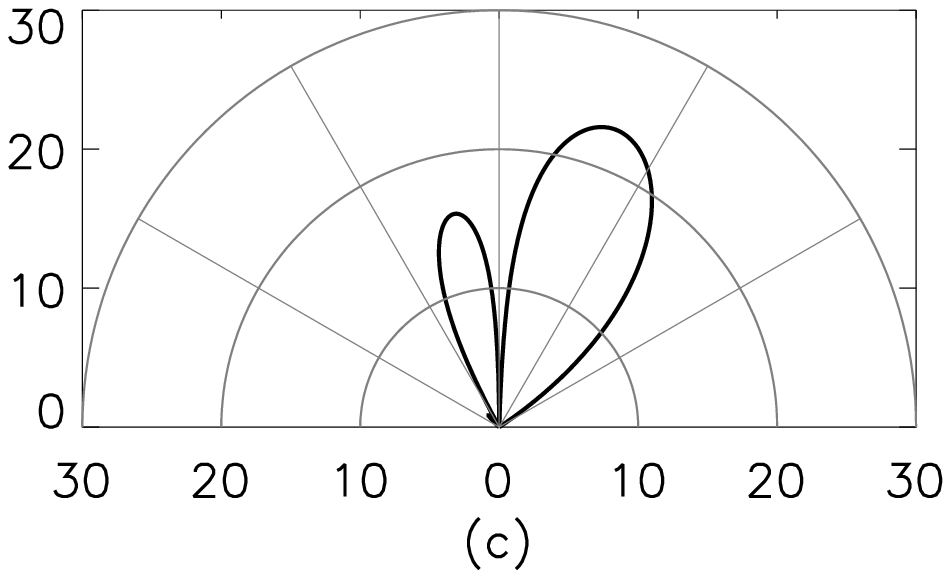}{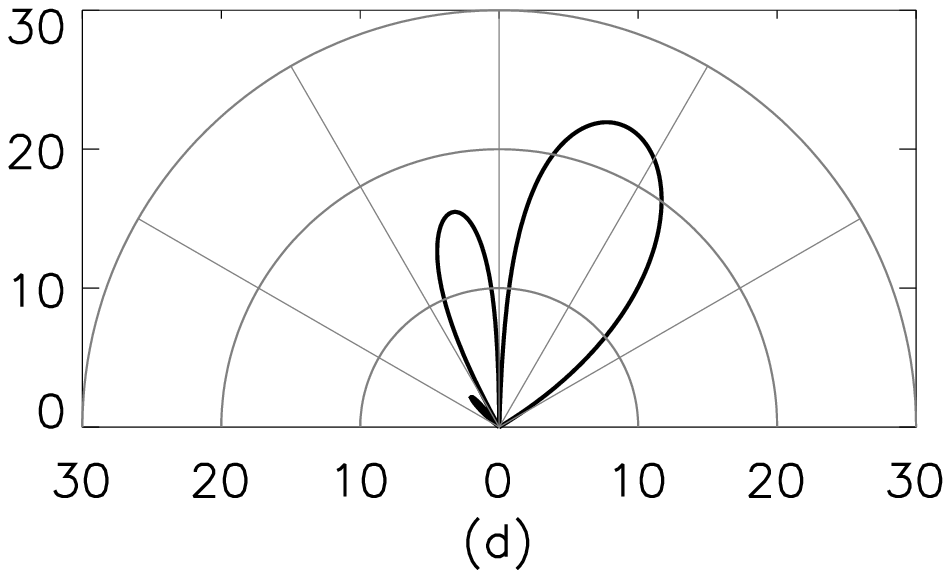}\\
\figcaption{West-east cuts through simulated MWA antenna tile patterns for a zenith (a,b) and a $28^\circ$ easterly pointing (c,d) at 150~MHz.  Panels (a) and (c) show the X polarization dipole power response pattern, while panels (b) and (d) show the Y polarization power response pattern.  The scale is logarithmic, with arbitrarily normalized decibel units.  The polarization-dependent gain structure is clearly visible in the sidelobes.  \label{fig:beams} }
\end{figure}

For present purposes, we assume the polarized primary beam patterns are identical across all tiles, and can be modeled
by simply summing together the direction-dependent complex gains of
the individual dipoles in the tile, i.e., mutual coupling between elements and tile-to-tile differences are ignored.  We model the
complex beam patterns of an isolated individual dipole for both the north-south (Y) and east-west (X) electric field polarizations using
the {\tt WIPL-D Pro}\footnote{http://www.wipl-d.com} electromagnetic
modeling software package.  A tile beam pattern is then computed by
summing the 16 dipole responses with the dipoles assumed to be at
their nominal locations in a tile and with the individual responses
modified by the nominal amplitudes and phases introduced by the
beamformer for the given delay settings.  Figure~\ref{fig:beams} shows
cuts through power patterns (square modulus of the complex beam) at
zenith and at a pointing direction $28^\circ$ east of zenith for both
the X and Y polarizations.  Model beam patterns were calculated at
frequency intervals of 2~MHz, since they vary significantly across the
MWA frequency band.

We assume that this model fully describes the direction dependence of
each tile.  We do, however, allow for a different overall, i.e.,
direction-independent, complex gain for each tile.  We follow the Jones
matrix formalism as presented by \citet{Hamaker1996}. The
instrumental model then takes the following form for a single tile at
a single frequency:
\begin{equation}
{\mathbf v}_{\rm A}={\mathbf G}_{\rm A} {\mathbf B} {\mathbf e}_{\rm A},
\end{equation}
where
\begin{equation}
{\mathbf e}_{\rm A} = \left(\begin{array}{c}e_{\rm x} \\ e_{\rm y} \end{array}\right),
\end{equation}
is the incident electric field at tile A, decomposed into linear E-W and N-S polarizations,
\begin{equation}
{\mathbf v}_{\rm A} = \left(\begin{array}{c}v_{\rm x} \\ v_{\rm y} \end{array}\right),
\end{equation}
is the vector of measured antenna voltages,
\begin{equation}
\label{eqn:gcal}
{\mathbf G}_{\rm A} = \left(\begin{array}{cc}g_{\rm A,x} & 0 \\ 0 & g_{\rm A,y} \end{array}\right),
\end{equation}
is the matrix of direction-independent complex gains for an antenna, and
\begin{equation}
\label{eqn:beam}
{\mathbf B} = \left(\begin{array}{cc} b_{\rm x}(\theta,\phi) & 0 \\ 0 &  b_{\rm y}(\theta,\phi)\end{array}\right),
\end{equation}
is the matrix of direction- and frequency-dependent but tile-independent gains due to the
primary beam shape (we represent spatial coordinates with $\theta$ and $\phi$). We neglect the feed-error ``D'' matrix of
\citet{Hamaker1996}; in other words we assume that the sensitivity of
the X-polarization response of a dipole to Y-polarization radiation
(the cross-polarization) is zero, and vice versa.  This is likely to
be a good approximation since ideal dipoles have zero
cross-polarization by definition.  In reality, various effects, such
as the finite thickness of our dipole elements, interactions between
structures in neighboring dipoles, or projection effects may produce a nonzero
cross-polarization response. In this paper we restrict our imaging and
analysis to these two senses of linear polarization and their
combination as total intensity. Errors caused by neglecting
cross-polarization effects are second order in the small off-diagonal
elements of the D-matrix.

The strategy we adopt for the data reduction is first to analyze short
snapshots wherein the settings in the analog beamformer were static so
that the primary beam pattern can be taken as constant over the duration of each snapshot, and any gain changes due to
the sidereal motion of the sky relative to the beam can be neglected.
In this regime, the direction-dependent gain can be factored out of
the response and corrected in the image plane in the resulting
map. Under this approximation, we are able to use standard tools
for radio astronomical data analysis for much of the processing.
Finally, the frequency dependences of the overall antenna complex gains
are in principle determined by modeling the summed spectra of the
bright sources in the field used in the calibration.  

Standard calibration procedures rely on being able to observe a field
containing a strong source with easily modeled structure that substantially dominates the visibilities.  For the two
fields presented in this work, Hydra~A is the strongest source in the
field. \citet{Lane2004} present low-frequency images that show that
while it is quite extended at the VLA's resolution, most of the flux
is contained within a region that is a few arcminutes in
radius. Since this extent is smaller than the angular resolution of
the MWA-32T, we were able to treat Hydra~A as a point source in our
calibration analysis.  One might expect that for a large field-of-view
instrument such as the MWA, we would also need to include several or
even many additional strong sources in the calibration model with
known direction-dependent gains.  We therefore experimented with
calibration models that included several point sources in addition to
Hydra~A, but we found that the complex tile gain solutions were not
significantly changed. We therefore simply used Hydra~A as the only
calibration source in subsequent analysis.  It should be noted that
this is a potential source of error.

\subsection{Ionosphere}

At the low radio frequencies of the MWA, position- and time-dependent
variations in the electron density of the ionosphere cause variations
in propagation times, which appear in the visibility data as
frequency- and time-dependent phase shifts.  For the short baselines
of the MWA-32T, these variations are, except at times of extreme
ionosphere disturbance, refractive in effect, i.e., they simply cause
apparent changes in the positions of point sources on the sky.  These
positions shifts may be different in different directions, especially
over a wide field such as that of the MWA, and consequently may lead to
distortions in the derived images.

Ionospheric effects have been quantified by studies with other
low-frequency interferometers.  \citet{Baldwin1985}, using the
Cambridge Low-Frequency Synthesis Telescope at 150 MHz, found that the
ionosphere typically caused $5^\circ$ (RMS) phase variations on 1~km
baselines on sub-day timescales, which were uncorrelated from
day to day. They also remarked that ionospheric irregularities on
large spatial scales, most likely related to the day-night cycle and
strongly correlated from one day to the next, could induce apparent
position shifts of sources of up to $20'$.  \citet{Kassim2007} found, with the VLA
operating at 74 MHz, that during times of moderate ionospheric
disturbance relative position variations on short timescales across a
$25^\circ$ field of view were at most $2'$.  Similarly,
\citet{Parsons2010}, using observations of bright sources with the
PAPER array at 150 MHz, found short-term small (typically less than
$1'$) position offsets that were not correlated from day to day, and
long-term large (up to $15'$) position offsets which were correlated
from day to day, and were mainly in the zenith direction.

In our analysis, we average snapshots taken with the center of the
field within several hours of the local meridian over a period of eight
days.  The results in the papers cited above suggest that uncorrelated
short-term variations in source directions will be significantly
smaller than our beam size of $\sim15'$ and, furthermore, that
they will tend to average out when images derived from individual
snapshots are combined.  We therefore neglect them.  Long-term
correlated variations in source directions may be comparable to our
beam size, and may not average to zero as we combine snapshot
images. However, our calibration strategy, described in
Section~\ref{sec:CalStrategy}, will tend to remove any ionospheric
offset at the position of Hydra~A through the phase terms in the
direction-independent gain solutions.  It is possible, depending on
the behavior of the ionosphere during the present observations, that long-term
differential position shifts of several arc minutes might be present
in our final images.  We investigate this possibility through comparisons of the
positions of our extracted sources with source positions listed in
published catalogues.

Our neglect of short-term ionospheric effects is justified only because of the small baselines ($<$350~m) of the MWA-32T array.  For the longer baselines ($\sim$3~km) of the full MWA, 
we believe that these effects will need to be corrected to achieve the dynamic range required for many of the science goals.

\section{Reduction Pipeline}
\label{sec:pipeline}

\subsection{Initial Processing and Editing}

We developed a calibration and imaging pipeline based
on the NRAO Common Astronomical Software
Applications\footnote{\tt{http://casa.nrao.edu}} (CASA) package and
additional tools that we developed in Python and IDL.  The pipeline uses a series of short observations to generate ``snapshot'' images which are weighted, combined, and jointly deconvolved to produce final integrated maps.

In the first
stage of the pipeline, the visibilities were averaged over 4-s
intervals and converted from the MWA instrumental format into 
UVFITS files.  The MWA-32T correlator does not perform fringe-stopping
(the correlation phase center is always at zenith), so phase rotations
were applied to the visibilities to track the desired phase center.
As a part of this process, data corrupted by RFI were flagged for
later exclusion from the analysis. The data were then imported into
CASA.  Additional editing was done to flag data affected by known
instrumental problems or RFI. Approximately $25\%$ of the data were
flagged at this stage, mainly because a problem in the data capture
software corrupted 480 of the 2080 correlation products.

\subsection{Calibration of Antenna Gains}

Calibration was performed separately for each snapshot with CASA,
using Hydra~A as the gain and phase reference (as discussed in
Section~\ref{sec:CalStrategy}).  Although Hydra~A was not at the
center of the primary beam during observations of the EoR2 field, it was still strong enough to
substantially dominate the visibilities.  Model visibilities were
calculated using a point source model for Hydra~A, assuming an unpolarized flux of unity.  The
overall frequency-dependent flux scale of the data was set at a later
point, along with a correction for the direction-dependent gain.
Time-independent channel-by-channel gain factors were calculated for
each tile using the task {\tt bandpass}.  After this was done and the
visibilities corrected for the gain factors, time-dependent overall
tile gain factors were calculated on a 32 second cadence using the
task {\tt gaincal}.  The factors determined in the two tasks give the
frequency- and time-dependent $g_{\rm A,x}$ and $g_{\rm A,y}$ terms of
Equation~\ref{eqn:gcal}.

The $g_{\rm A,x}$ and $g_{\rm A,y}$ terms were examined for temporal stability and
spectral smoothness; regions where deviations were apparent were
flagged. Such deviations were rare, and an important outcome of this
analysis is the recognition that the MWA antenna gains are quite
stable over frequency and time.  In fact, the gains even tended to be
stable from one day to the next.  However, some complicated variations
in gain as a function of frequency were identified.  These were
associated with damaged cables and connectors that have since been
replaced or repaired.


\subsection{Snapshot Imaging}

The data from each snapshot were subdivided into 7.68~MHz wide
frequency bands, and multifrequency synthesis imaging was performed for each snapshot using the {\tt CASA} task
{\tt clean}.  Images were made with a $3'$ cell size over a
$\sim51^\circ \times \sim51^\circ$ patch of sky in order to cover
the majority of the main lobe of the primary beam. The ``XX'' and
``YY'' polarizations were imaged separately. Conversion to the
standard Stokes parameters was not performed at this stage, since, as
discussed in Section~\ref{sec:CalStrategy}, the gains for the two polarizations have
different direction dependence.  The ``w-projection''
algorithm \citep{Cornwell2008} was used to correct for wide
field-of-view effects, and to produce an image with an approximately
invariant point spread function in each of the snapshot images.  The
images were deconvolved using the Cotton-Schwab {\tt CLEAN} algorithm
(see \citealt{Schwab1984}) down to a threshold of 1\% of the peak flux
in the image.

A position-dependent ``noise'' map was also computed for each of the 7.68~MHz wide snapshot images by selecting a 64-pixel by 64-pixel
window around each pixel in the image and fitting a Gaussian to the central
80\% of a histogram of the pixel values.  This procedure was employed
because of the high point-source density in these maps.  Throughout
much of the area of these maps, it was impossible to identify a source
free region from which to estimate the background noise fluctuations,
and the presence of sources artificially skewed the noise estimates
calculated strictly as the root-mean-square (RMS) of
the pixel values.  We found that this clipped histogram fitting
procedure provided a more robust estimate of the RMS of the background
noise.  These noise maps were smoothed on a $1^\circ$ scale to remove
local anomalies introduced by extended or clustered sources.  However, despite
these procedures, some areas, particularly in especially crowded
regions, still had anomalously high noise estimates.

As discussed above, each snapshot was only $\sim$5 minutes in duration, and was obtained while the delay line
settings in the analog beamformers were fixed.  This allowed us to model the primary beam
pattern of each tile as fixed relative to the sky for the duration of
the snapshot. Our calculated model beams for each polarization formed
the frequency-dependent $b_x(\theta,\phi)$ and $b_y(\theta,\phi)$
terms of Equation~\ref{eqn:beam}.  These terms are time-dependent only
in that they are different for each snapshot.

\subsection{Snapshot Combination and Joint Deconvolution \label{sec:weighting}}

Deeper images were obtained by combining snapshot
maps from a particular 7.68 MHz wide band according to:
\begin{equation}
I_{\rm dirty}(\theta,\phi)=\frac{ \sum_i \frac{D_i(\theta,\phi) B_i(\theta,\phi)}{\sigma_i^2(\theta,\phi)}}{\sum_i \frac{B_i^2(\theta,\phi)}{\sigma_i^2(\theta,\phi)}},
\end{equation} 
where $I_{\rm dirty}$ is the integrated, primary-beam corrected, dirty
map, the snapshots are distinguished by index $i$, $D_i$ is a snapshot
dirty map, $B_i$ is the primary beam calculated for that snapshot, and
$\sigma_i$ is the fitted rms noise obtained from the noise map for
snapshot $i$.  This combination optimizes the signal to noise ratio of
the final image.  Beam patterns are calculated independently for each 7.68~MHz channel.  The same weighting scheme was used to combine the ``clean components'' and
residual maps of the individual snapshots.

A variant of the H\"ogbom CLEAN algorithm \citep{Hogbom1974} was used to further deconvolve the integrated residual maps, using a position-dependent point spread function calculated using the same weighting scheme:
\begin{equation}
P_{{\rm int},j}(\theta,\phi)=\sum_i \frac{B_i(\theta,\phi) B_i(\theta_j,\phi_j) P_i(\theta-\theta_j,\phi-\phi_j)}{\sigma_i^2(\theta,\phi)},
\end{equation}
\noindent where $P_{{\rm int},j}$ is the point spread function for a source at position $(\theta_j,\phi_j)$, $B_i$ is the primary beam pattern for the $i{\rm th}$ snapshot with a PSF given by $P_i$, and the peak of the function is normalized to unity.  CLEAN components were selected by choosing the pixel in the residual map with the largest signal-to-noise ratio (SNR, determined by dividing the residual map by its noise map).  The PSF was scaled to a peak of 10\% of the flux of the pixel.  The images were restored with a Gaussian beam determined by a fit to the weighted average of the individual snapshot point spread functions at the field center.

\subsection{Averaging}

In order to increase the signal to noise ratio and image fidelity for source detection
and characterization, the individual 7.68~MHz maps, after deconvolution and restoration, were averaged
together. An approximate flux scale for the maps was first set by scaling the surface brightness of the maximum pixel at the location of Hydra A to a value of $296 \times (\nu/150~{\rm MHz})^{-0.91}~{\rm Jy}~{\rm beam}^{-1}$ (this model was derived from fitting a power law to other low-frequency measurements).  For each field, 30.72~MHz bandwidth maps centered at
123.52~MHz, 154.24~MHz, and 184.96~MHz were each made from four
7.68-MHz maps. Before averaging, the 7.68-MHz map fluxes were scaled
to the averaged map frequency using an assumed spectral index of $\alpha=-0.8$ (where $S \propto \nu^\alpha$).  The averages were computed in a weighted sense using the integrated primary beam weights from each map.  A full-band (92.16 MHz bandwidth) weighted average map was made from the three
30.72-MHz bandwidth maps after scaling them to a common reference
frequency of 154.24~MHz, again using a spectral index of $\alpha=-0.8$.
The portion of each field within $25^\circ$ of the field center was
used for the subsequent analysis.

\subsection{Source Extraction\label{sec:extraction}}

Sources were identified in each full-band, i.e., 92.16 MHz wide,
averaged map using an automated source extraction pipeline.  The
first step of this pipeline was the calculation of a
position-dependent noise map using the method described in Section
\ref{sec:weighting}.  The full-band map was then divided by the noise
map to produce a SNR map.  An iterative
process then was initiated by identifying contiguous regions of pixels
above a certain SNR detection threshold and, for each such region, defining a
fitting region that extended beyond the set of connected pixels by
several synthesized beamwidths.  A two-dimensional Gaussian fit was performed on the
corresponding region in the full-band map.  The parameters determined
in each fit included the background level, peak position, peak
amplitude, major axis width, minor axis width, and position angle. If
the fit converged to a Gaussian centered within the fitting region,
then the source was subtracted from the map, and the extracted source
parameters were recorded.  After fitting all regions identified for a
certain SNR level, the detection threshold was reduced and the process was
repeated.  Regions above the detection threshold for which a fit failed to
converge are refit in subsequent iterations at lower detection thresholds
(where the fitting regions are typically larger in size).  For
completeness, sources were extracted down to a detection SNR threshold of 3.  It
should be noted that in this fitting procedure, each region that is fit by a single two-dimensional Gaussian is taken to correspond to a separate source.  Sources that are too close together to be resolved into separate components will be fit with a single component and erroneously taken to be a single source that is a ``blend'' of the two components.

Sources that were identified in the full-band average map were then
extracted from each of the 30.72-MHz bandwidth maps.  The sources were
sorted by their detection signal-to-noise level and, for each of the three maps, were fitted in
descending order.  For each source, the position and shape (axial
ratio and position angle) parameters of the Gaussian fitting function
were held fixed to the values determined in the full-band map
extraction, while the peak value, background level, and a scaling
factor for the widths of both the major and minor axes of the Gaussian were
allowed to vary.  This procedure was performed for all sources.  When
the fit successfully converged, the best-fit model was subtracted from
the sub-band map.  A total of 908 sources were extracted
in the Hydra~A field and 1100 sources were extracted from the EoR2
field.

\subsection{Astrometric Corrections \& Flux Calibration\label{sec:astfluxcal}}

We compared the positions and relative fluxes of the sources
identified in the full-band maps with the positions and fluxes of
possible counterparts in other catalogs.  These comparisons formed the
basis for astrometric corrections, and for the determination of the
overall MWA flux scale.  Counterparts to MWA-32T sources were identified at
408~MHz by locating sources in the MRC which were within $15'$
of an MWA source.  Although we expect our astrometric accuracy to be
much better than $15'$, this value was chosen to be comparable
to the size of the MWA-32T synthesized beam major axis full width at
half maximum response, viz., $13'$ for the
Hydra~A field and $14'$ for the EoR2 field, and to allow for
some degree of systematic error in the MWA source positions.  To avoid
possible blending issues, we only considered an identification to be
secure when there was precisely one source in the MRC within
$30'$ of the (pre-adjustment) MWA source position. A total of
419 sources were matched uniquely to the MRC in the Hydra~A field and
520 sources were matched in the EoR2 field.

\begin{figure}
\plotone{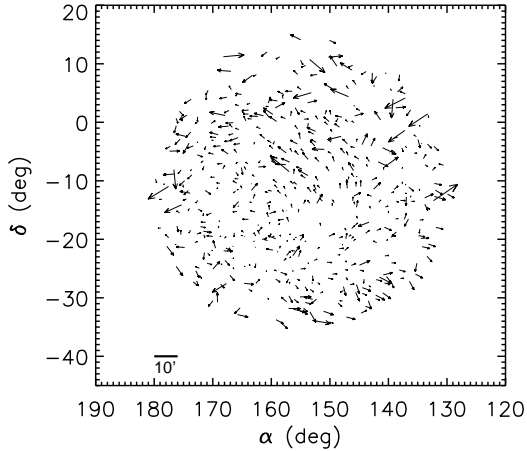}
\figcaption[position_vectorplot.eps]{Spatial offsets between the positions of MWA sources in the EoR2 field and matched sources in the Molonglo Reference Catalog \citep{Large1981}.  An overall coordinate system shift has been removed.\label{fig:position_vectorplot}}
\end{figure}

\begin{figure}
\includegraphics[width=0.48\linewidth]{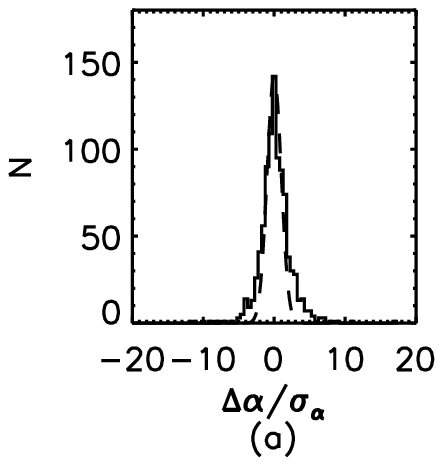}
\includegraphics[width=0.48\linewidth]{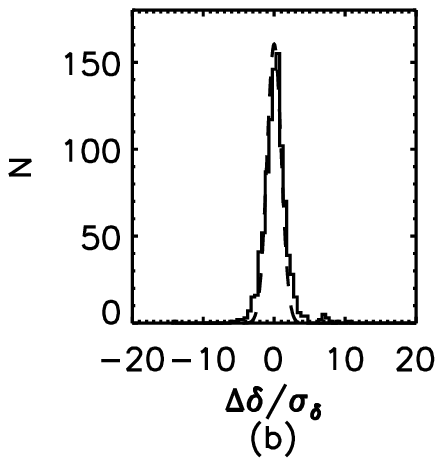}
\figcaption[position_histograms.eps]{Histograms of the normalized right ascension, $\alpha$, (a) and declination, $\delta$, (b) errors for the extracted MWA sources relative to matched sources in the Molonglo Reference Catalog \citep{Large1981}.  The standard deviations, $\sigma_\alpha$, and $\sigma_\delta$, are calculated following \citet{Condon1997} with the simplifying assumption of circular source geometry  Assuming Gaussian error properties, the residual distribution should approximate a standard normal distribution, which is over-plotted with a dashed line.  Hydra~A is omitted from the histograms.\label{fig:position_hist}}
\end{figure}

An astrometric correction was then calculated by allowing for a linear
transformation of the MWA source coordinates in order to minimize the
positional differences between corresponding MWA and MRC sources.  The
transformation permits offsets in both right ascension and declination
as well as rotation and shear with respect to the field
center. Optimal transformation parameters were determined by
performing weighted least-squares fits. The results from initial fits
indicated that there were errors in the (Earth-referenced) coordinates
of the MWA tiles and in the conversion of coordinates in the maps from
the epoch of observation frame to the J2000 frame. These errors were
then corrected.  The final fits were found to be consistent solely
with offsets of the MWA source coordinates, with no shear or rotational
effects.  We therefore applied offsets of $\Delta\alpha = -0.6'$
and $\Delta\delta = 1.6'$ to the positions of the sources in the
Hydra~A field, and of $\Delta\alpha = 2.2'$ and $\Delta\delta =
1.9'$ to those in the EoR2 field. We believe the coordinate
offsets are likely due to a combination of the effects of structure in
Hydra~A and to ionospheric refraction.
Figure~\ref{fig:position_vectorplot} shows the post-offset-correction
differences in position between corresponding MWA and MRC sources in
the EoR2 field. Histograms of residual positional differences from both the
Hydra~A and EoR2 fields are plotted in Figure~\ref{fig:position_hist}.
For this figure the difference for each source is normalized by the
expected error based on the signal to noise ratio for the intensity
and the 32T synthesized beamwidth (with the assumption of a circular
source, see \citealt{Condon1997}).  The residual position differences
are generally consistent with those expected, even though there are a
small number of large position differences.

\begin{figure}
\centering
\plotone{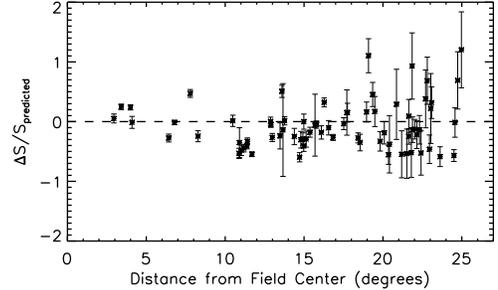}
\figcaption{The fractional differences between predicted and measured fluxes for the EoR2 field at 154.24~MHz (described in Section~\ref{sec:astfluxcal}), where $\Delta$S is defined as the MWA-32T measured flux minus the flux predicted from fitting MRC and Culgoora measurements. The error bars are derived from combining the RMS noise in the MWA map at the source position in quadrature with the flux prediction uncertainty.  The differences are displayed as a function of distance from the field center, in order to assess the presence of any radial biases in the MWA flux measurements.  No significant bias was found out to large distances from the field center. \label{fig:radial_flux_diff}}
\end{figure}

A final flux scale was set for each map using the MWA sources which had counterparts in both the MRC and Culgoora source lists.  Only sources in the MWA catalog above a detection SNR threshold of 5 were used. Hydra~A was excluded for reasons discussed in Section~\ref{sec:results}. A prediction for each source was obtained by fitting a power law spectrum to the 408~MHz MRC flux and the 80~MHz and 160~MHz Culgoora fluxes.  Under these criteria, measurements in all 3 bands were found for a total of 64 uniquely matching sources in the Hydra~A field and 81 uniquely matching sources in the EoR2 field.  Using these flux predictions, a flux scale correction was calculated of the form:
\begin{equation}
S_{\rm calibrated}=C_{\nu} \times S_{\rm uncalibrated}.
\end{equation}
These calibration terms were calculated independently for each of the three sub-band maps as well as for each averaged map for each field.  For the Hydra~A field, the calculation yeilded $C_{\nu} = 1.26$ for the full-band average map, 1.17 for the 123.52~MHz map, 1.20 for the 154.24~MHz map and 1.18 for the 184.96~MHz map, and for the EoR2 field, the calculation yielded $C_{\nu}=1.24$ for the full-band average map, 1.19 for the 123.52~MHz map, 1.23 for the 154.24~MHz map and 1.17 for the 184.96~MHz map.  The residuals after applying these flux scale corrections were analyzed to determine if additional biases were present as a function of position in the image (biases would potentially be seen, e.g., if the assumed primary beam model was incorrect).  An example of these plots is shown in Figure~\ref{fig:radial_flux_diff}, which shows no evidence for a radially increasing flux bias. 

The magnitude of the post-correction residual differences between the MWA measurements and the predicted fluxes for calibration sources are still larger on average than expected under the assumption that the uncertainty in each MWA-32T flux measurement is due to the RMS map noise at the source location, and that the uncertainty in the predicted flux of each source is propagated for the power-law fitting procedure.  This excess in the differences could be due to errors in the spectral model for the calibration sources, temporal variability of the sources, or to as yet unidentified errors.  Since this is the first work based on
MWA-32T data to report the fluxes of a large number of sources, we make the conservative assumption that these excessively large residuals are due solely to errors in the MWA-32T flux measurements.  We assume the flux uncertainties follow a Gaussian distribution which includes the effects of the RMS map noise added in quadrature with an additional component proportional to the measured flux of the source:
\begin{equation}
\label{eqn:uncertainty}
\sigma_{\rm MWA}^2 = \beta^2 S_{\rm MWA}^2 + \sigma_{\rm Map}^2,
\end{equation}
where $\sigma_{\rm MWA}$ is the $1\sigma$ flux uncertainty for a particular source, $\beta$ is the fractional flux uncertainty, $S_{\rm MWA}$ is the measured source flux, and $\sigma_{\rm Map}$ is the RMS map noise at the position of the source.  We evaluated the standard deviation, $\sigma_D$, of the fractional flux difference, $D$, where $D$ is calculated as:
\begin{equation}
D=\frac{S_{\rm MWA} - S_{\rm Predicted}}{S_{\rm Predicted}}.
\end{equation}
We then solved for the fractional uncertainty in the MWA measurements which would be needed to account for the magnitude of the measured value of $\sigma_D$:
\begin{equation}
\beta^2=\sigma_D^2 \left(\frac{S_{\rm Predicted}}{S_{\rm MWA}}\right)^2 - \left(\frac{\sigma_{\rm Predicted}}{S_{\rm Predicted}}\right)^2 - \left(\frac{\sigma_{\rm Map}}{S_{\rm MWA}}\right)^2.
\end{equation}
We calculated the average value of $\beta$ separately for each field.  For the higher-frequency maps, we found that the values of $\beta$ were much larger far from the field center where the primary beam approaches the first null; for these maps the sources were separated into inner and outer region sets using a cutoff of $18^\circ$, and $\beta$ was calculated separately for each region.  Using these results, we assign fractional flux uncertainties of 30\% for all sources in the full-band average maps, 35\% for all sources in the 123.52~MHz maps, 35\% for sources in the inner region of the 154.24~MHz maps, 60\% for sources in the outer region of the 154.24~MHz maps, 35\% for sources in the inner region of the 184.96~MHz maps and 80\% for sources in the outer region of the 184.96~MHz maps.  These fractional uncertainty values are applied to all sources in the catalog by adding them in quadrature to the map RMS values as described in Equation~\ref{eqn:uncertainty}.


\section{Results}

\subsection{Radio Maps and Source Catalog\label{sec:results}}

\begin{figure*}
\plotone{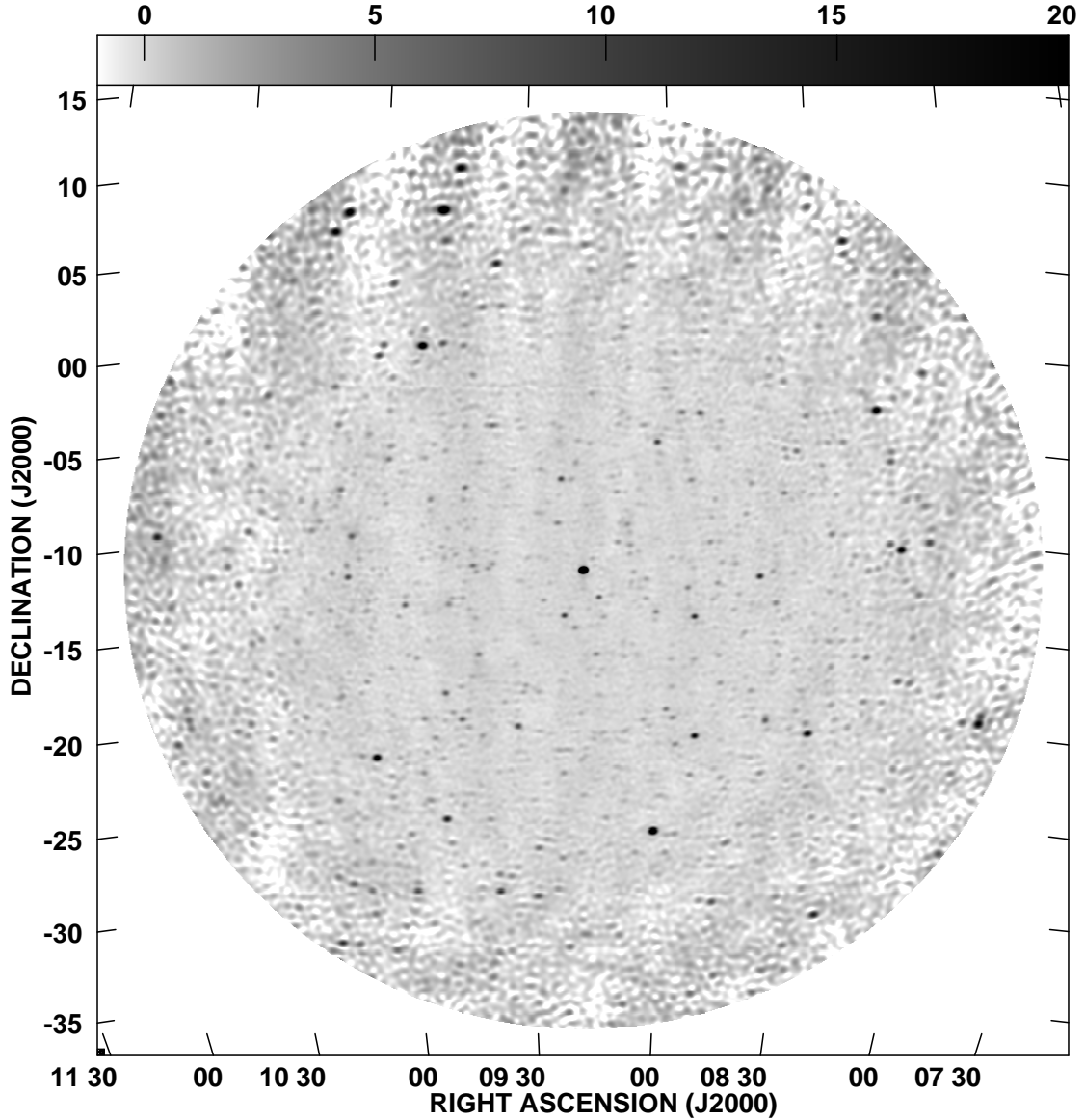}
\caption{Full bandwidth synthesis map of the Hydra~A field, with a $50^\circ$ diameter.  This map was produced using the pipeline described in Section~\ref{sec:pipeline}, and served as a basis for source identification in this field.\label{fig:HydAmap}}
\end{figure*}

\begin{figure*}
\plotone{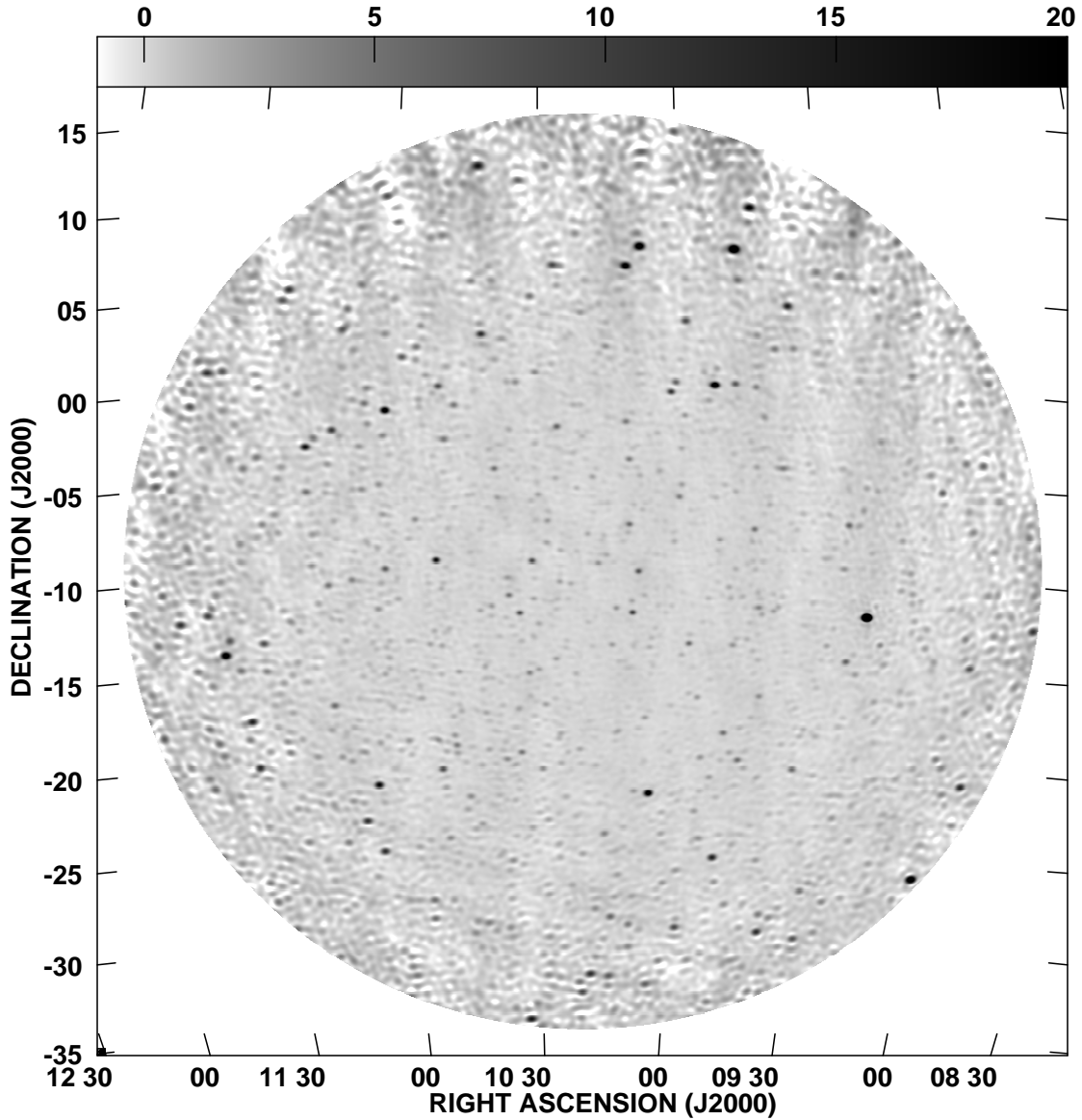}
\caption{Full bandwidth synthesis maps of the EoR2 field, with a $50^\circ$ diameter.  This map was produced using the pipeline described in Section~\ref{sec:pipeline}, and served as a basis for source identification in this field.\label{fig:EoR2map}}
\end{figure*}

\begin{figure}
\includegraphics[width=0.48\linewidth]{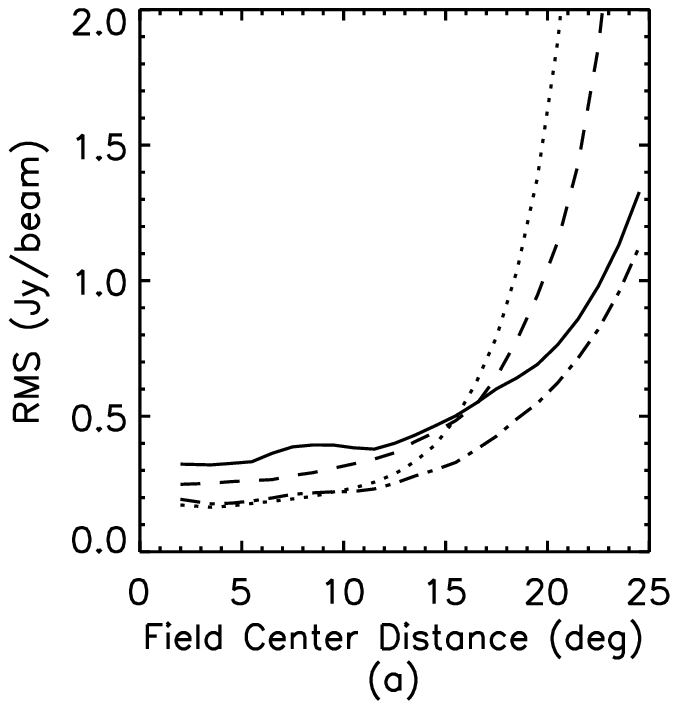}
\includegraphics[width=0.48\linewidth]{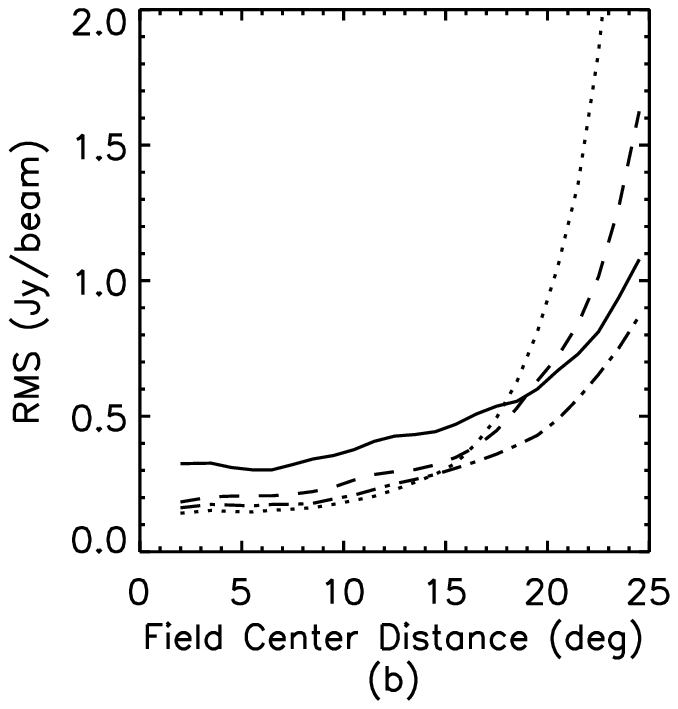}
\caption{Radial dependence of the calculated RMS noise in the Hydra~A (a) and EoR2 (b) field images.  The RMS is plotted for the three sub-band maps, with central frequencies of 123.52~MHz (solid lines), 154.24~MHz (dashed lines), and 184.96~MHz (dotted line) as well as for the full-band averaged maps (dot-dashed lines).  The values shown are the medians of $1^\circ$ wide radial annuli.  The points are connected with lines for clarity.  The frequency dependent primary beam shape is evident from the increasing RMS at large distance from the field center, and the confusion limited nature of the maps is illustrated by the flat RMS profile near the center of the field.  The minimum RMS approaches a value of $\sim$130~mJy/beam.\label{fig:map_rms}}
\end{figure}

The full band average maps of the Hydra~A and EoR2 fields are
displayed in Figure~\ref{fig:HydAmap} and Figure~\ref{fig:EoR2map}.
These images overlap partially.  Together, they cover $\sim$2700 square
degrees.  The synthesized beam for the Hydra~A field  has a major-axis
width of $19'$ in the 124.52~MHz map, $14'$ in the 154.24~MHz map, $12'$ in the 184.96~MHz map, and $13'$ in the full band average map.  For the EoR2 field, the major-axis beam widths are $18'$ in the 124.52~MHz map, $16'$ in the 154.24~MHz map, $13'$ in the 184.96~MHz map, and $14'$ in the full band average map.

The bright radio galaxy Hydra~A is the dominant source in these maps.  The flux of the source is measured to be $710\pm210~{\rm Jy}$ in the full band average map of the Hydra~A field, and $550\pm170~{\rm Jy}$ in the full band average map of the EoR2 field.  These measurements of Hydra~A are significantly brighter than expected based on previous measurements: the Culgoora 160~MHz measurements give a flux of 243~Jy, and a prediction based off of the Culgoora and MRC measurements (as described in Section~\ref{sec:astfluxcal}) gives a flux of 284~Jy.  Although Hydra~A is slightly extended in the MWA-32T maps, the structures seen in the previous low frequency maps of Hydra~A presented by \citet{Lane2004} are below the scale of the MWA synthesized beam.  We note that the MWA-32T array has a significantly more compact $uv$ distribution than the VLA, Molonglo or Culgoora telescopes.  A flux from Hydra~A above what is expected from the Culgoora measurements is also noted in measurements with the PAPER array (\citealt{Jacobs2011pc}, private communication), which has a similar baseline distribution to that of the MWA-32T.  Consequently we did not include Hydra~A in the final flux scale calibration procedure described in Section~\ref{sec:astfluxcal}.

The behavior of the fitted RMS noise in these images is illustrated in Figure~\ref{fig:map_rms}, which shows annular averages as a function of distance from the field center.    We can estimate a lower limit to the RMS noise in each map by calculating the classical source confusion limit.  Note that this differs from the sidelobe confusion limit; see, e.g. \citealt{Condon1974} for a rigorous discussion of classical source confusion in radio telescopes.  \citet{DiMatteo2002} present a model of the radio source counts derived from the 6C \citep{Hales1988} catalog at 150~MHz.  Their expression takes the form of a broken power law\footnote{The notation of \citet{DiMatteo2002} is ambiguous, and is interpreted differently by several authors.  We note that \citet{Lidz2008} quote a modified form of the expression, which affects the normalization of the power law.  Although the \citet{Lidz2008} expression fits the 6C source counts slightly better at high flux values, the formulation presented in this paper fits the data better throughout the entire flux range.  The expression from \citet{DiMatteo2002} has been corrected with an additional minus sign to make it continuous across the transition at $S=S_0$.}:
\begin{equation}
\scriptsize
\frac{dn}{dS} = \begin{cases} 
4000~(\frac{S}{1~{\rm Jy}})^{-2.51}~{\rm sources}~{\rm Jy}^{-1}~{\rm sr}^{-1},  & S > S_0\\
4000~(\frac{S_0}{1~{\rm Jy}})^{-0.76}~(\frac{S}{1~{\rm Jy}})^{-1.75}~{\rm sources}~{\rm Jy}^{-1}~{\rm sr}^{-1},  & S < S_0
\end{cases}
\end{equation}
where $S_0=0.88~{\rm Jy}$.  Integration of this expression gives the source density in each beam above a minimum flux value, $S_{\rm min}$:  
\begin{equation}
\rho_S=\frac{\pi \theta^2}{4 \ln{2}} \int_{S_{\rm min}}^{\infty}\frac{dn}{dS} dS~{\rm sources}~{\rm beam}^{-1},
\end{equation}
where $\rho_S$ is the source density in units of sources per beam, and $\theta$ is the FWHM synthesized beam size.
The source confusion limit then corresponds to the flux, $S_{\rm min}$, for which the source density approaches one source per synthesized beam area (typically maps are considered to be source confusion limited when they have a source density of greater than 1 source per 10 synthesized beams).  The average RMS noise (Figure~\ref{fig:map_rms}) reaches a minimum value of $\sim$160~mJy for the full band average map of the EoR2 field.  Using the corresponding flux threshold for an unresolved source of $S_{\rm min}=160~{\rm mJy}$ in combination with the EoR2 full band map synthesized beam area gives $\rho_S \sim 0.30$ sources per synthesized beam, or, in other words, one source of at least 160~mJy in roughly every three synthesized beams.   Estimates for the other average and sub-band maps give expected source densities of one source per every 3 to 6 synthesized beams .  Thus, source confusion is likely the limiting source of noise in the central region of these maps.  This explains the relatively flat nature of the central noise floor seen in Figure~\ref{fig:map_rms}.  Near the edges of the images, not far from the first null of the primary beam, we expect the noise to be dominated by receiver noise, and to scale with the inverse of the primary beam power pattern. This is also seen in Figure~\ref{fig:map_rms} (the frequency dependence of the noise curves illustrates the frequency dependence of the primary beam).

A catalog was constructed from the 2008 detections at a detection SNR threshold $\geq3$ of
potential sources in the EoR2 and Hydra~A fields. In the cases where
there were detections at corresponding celestial positions in the two
fields, the measurement where the source was closer to the observation
field center was retained, resulting in a list of 1526 unique source
detections.  The quality of this source list is assessed in
Section~\ref{sec:reliability} and Section~\ref{sec:completeness} as a
function of the detection SNR level.  The 655 sources detected at an SNR level $\geq$5 in the
detection images are reported in Table~\ref{tab:sources}.  The 871 sources with 3$\leq$SNR$<$5 are
considered to be less reliable detections.  A list of these candidate detections can be obtained by contacting the
authors.

\begin{deluxetable*}{cccccccccc}
\tablecaption{Detected sources in the the Hydra~A and EOR~2 fields}

\tablehead{
\colhead{Name} & \colhead{RA} & \colhead{DEC} & \colhead{$S_{\rm avg}$}& \colhead{$S_{123.52}$} & \colhead{$S_{154.24}$} & \colhead{$S_{184.96}$} & \colhead{Field} & \colhead{$r_{\rm FC}$} & \colhead{Detection}\\
& & & & & & & & & \colhead{SNR Level}
}
\tabletypesize{\scriptsize}
\startdata
J0747$-$1854 & 07h47m05s & $-18^\circ54'12''$ & 8.8$\pm$2.8 & 10.1$\pm$3.7 & 9.2$\pm$6.0 & 8.9$\pm$8.5 & HydA & $22.9^\circ$ &    8.6\\
J0747$-$1919 & 07h47m27s & $-19^\circ19'33''$ & 22.2$\pm$6.7 & 25.2$\pm$8.9 & 23.8$\pm$14.5 & 15.1$\pm$13.0 & HydA & $23.0^\circ$ &   22.4\\
J0751$-$1919 & 07h51m20s & $-19^\circ19'30''$ & 7.1$\pm$2.2 & 8.1$\pm$3.0 & 8.5$\pm$5.3 & 4.7$\pm$5.1 & HydA & $22.1^\circ$ &    8.7\\
J0752$-$2204 & 07h52m30s & $-22^\circ04'38''$ & 4.4$\pm$1.5 & 4.7$\pm$1.8 & 7.1$\pm$4.5 & 14.4$\pm$12.1 & HydA & $22.7^\circ$ &    5.6\\
J0752$-$2627 & 07h52m30s & $-26^\circ27'43''$ & 8.4$\pm$2.7 & 9.8$\pm$3.7 & 3.6$\pm$3.1 & \nodata & HydA & $24.7^\circ$ &    7.0\\
J0757$-$1137 & 07h57m10s & $-11^\circ37'10''$ & 3.9$\pm$1.3 & 4.7$\pm$1.8 & 5.1$\pm$3.3 & 2.5$\pm$2.5 & HydA & $19.8^\circ$ &    6.0\\
J0802$-$0915 & 08h02m18s & $-09^\circ15'32''$ & 3.4$\pm$1.2 & 5.8$\pm$2.2 & 2.1$\pm$1.5 & 1.9$\pm$1.8 & HydA & $18.8^\circ$ &    5.0\\
J0802$-$0958 & 08h02m34s & $-09^\circ58'55''$ & 8.9$\pm$2.8 & 12.4$\pm$4.4 & 6.0$\pm$3.7 & 9.1$\pm$7.3 & HydA & $18.6^\circ$ &   12.6\\
J0803$-$0804 & 08h03m60s & $-08^\circ04'48''$ & 4.5$\pm$1.4 & 6.5$\pm$2.4 & 2.4$\pm$1.6 & 6.7$\pm$5.4 & HydA & $18.7^\circ$ &    7.0\\
J0804$-$1244 & 08h04m17s & $-12^\circ44'32''$ & 4.6$\pm$1.5 & 4.8$\pm$1.8 & 5.2$\pm$3.2 & 4.5$\pm$3.7 & HydA & $18.0^\circ$ &    9.3\\
J0804$-$1726 & 08h04m42s & $-17^\circ26'41''$ & 4.9$\pm$1.5 & 5.8$\pm$2.1 & 4.5$\pm$2.8 & 3.1$\pm$2.6 & HydA & $18.5^\circ$ &    9.1\\
J0804$-$1502 & 08h04m53s & $-15^\circ02'54''$ & 2.8$\pm$0.9 & 2.2$\pm$1.0 & 4.6$\pm$2.9 & 1.0$\pm$1.1 & HydA & $18.0^\circ$ &    6.4\\
J0805$-$0100 & 08h05m30s & $-01^\circ00'12''$ & 9.4$\pm$2.9 & 10.8$\pm$3.9 & 8.3$\pm$5.2 & 6.4$\pm$5.4 & HydA & $21.1^\circ$ &   10.4\\
J0805$-$0739 & 08h05m40s & $-07^\circ39'22''$ & 3.5$\pm$1.2 & 4.4$\pm$1.7 & 3.7$\pm$2.4 & 1.9$\pm$1.9 & HydA & $18.4^\circ$ &    5.5\\
J0806$-$2204 & 08h06m26s & $-22^\circ04'43''$ & 3.7$\pm$1.2 & 4.2$\pm$1.6 & 3.8$\pm$2.4 & 3.6$\pm$3.1 & HydA & $19.8^\circ$ &    6.2\\
\enddata

\tablecomments{
The flux of each source detected in the MWA full-band averaged maps is presented along with the flux measured in each 30.72~MHz sub-band.  Duplicate sources in the region of overlap of the two fields are not listed.  Missing data indicates that the automatic source measurement algorithm failed to converge in a flux fit for that source in the sub-band map.  The field from which each source measurement comes from is listed, along with the distance of the source from the center of the field ($r_{\rm FC}$).  We expect systematic errors to be larger for sources far from the field center.  The ``Detection SNR Level'' indicates the signal-to-noise ratio at which the source was detected in the full-band averaged map.  Section~\ref{sec:reliability} discusses the reliability of the catalog at different detection SNR levels.  This list includes sources identified above a detection SNR threshold of 5.  The full source list of all sources above a detection SNR threshold of 3 is available from the authors.\\
\\
Table~\ref{tab:sources} is published in its entirety in the electronic edition of the Astrophysical Journal. A sample is shown here to illustrate its content.}

\label{tab:sources}
\end{deluxetable*}

\subsection{Reliability of the Source List\label{sec:reliability}}

The reliability of the identified sources was
evaluated through comparison with the flux limited sample from the
MRC, the VLSS, and maps from the TGSS.  The MRC source list has a well-defined completeness flux
limit and covers our entire field; however it gives fluxes at a
different frequency (408~MHz) and does not go quite as deep as our
survey.  The VLSS covers a portion of our fields at a lower frequency than the MWA (74~MHz), and provides a useful complementary assessment.
The TGSS maps are at the same frequency as the MWA
observations (150~MHz), but the maps which have been released to
date only cover a small fraction of our fields and are based on a
significantly different sampling of the visibility function due to
the different array baselines.

In order to assess the MWA-32T catalog reliability, we first evaluate the detectability of an MWA source in the external comparison survey.  The MWA full band average flux is extrapolated to the relevant frequency using a spectral index of $\alpha=-0.8$, and the extrapolated value is then compared to the parameters of the comparison map or catalog to evaluate whether it meets the detection criteria for that survey.  If the source is deemed detectable, then we search for a companion source in that catalog or map to see if the source was actually detected by the other survey.  Under the assumption that the other surveys are complete, this allows us to assess how many spurious sources are present in the MWA catalog.  We define the reliability as:
\begin{equation}
R=\frac{N_{\rm detected} / N_{\rm detectable} -f}{1-f},
\end{equation}
where $R$ is the fraction of MWA sources we believe to be reliable, $N_{\rm detectable}$ is the number of MWA sources which we believe should have been detectable in the comparison survey, $N_{\rm detected}$ is the number of detectable MWA sources for which we found counterparts in the other survey, and $f$ is the false source coincidence fraction.  We determine $f$ by calculating the source density of the comparison survey in the MWA fields, and use our counterpart matching criteria to estimate the probability that a randomly chosen sky location will lead to an association with a source in the comparison survey.  This analysis was performed for different MWA source detection thresholds.

For the reliability comparison with the MRC, we used the completeness limit of 1~Jy \citep{Large1981} to assess the detectability of the extrapolated MWA source fluxes.  An MRC counterpart is associated with the MWA source if it is within $10'$ of the MWA source position.  Based off of the counterpart search radius and source densities in the MRC field, we estimate a false coincidence chance of  4\%.  A fixed flux completeness limit is not given for the VLSS, however \citet{Cohen2007} note that for a typical VLSS RMS of $0.1~{\rm Jy}~{\rm beam}^{-1}$, the 50\% point source detection limit is approximately $0.7~{\rm Jy}$.  
We then assume the VLSS is complete down to a flux level of 1~Jy, and we again use a $10'$ source association radius in the reliability calculation.  At the present time, the VLSS catalog does not cover the entire combined EoR2 and Hydra~A region that we have surveyed.  To ensure we are only including sources in the VLSS survey area, we only analyzed MWA sources above a declination of $\delta=-25^\circ$.  We estimate a false coincidence chance of 18\% for the VLSS matching.

The cumulative and differential catalog reliabilities are listed as functions of the MWA detection level in Table~\ref{tab:cum_reliability} and Table~\ref{tab:dif_reliability}.  We view these reliability estimates
as lower limits, particularly at the lower flux levels, because our
assessments of comparison survey detectability do not take into account errors in
the source flux extrapolation or source time variability.  MWA sources
which are erroneously calculated as detectable will not, in general,
lead to detections of counterparts in the comparison catalog, whereas sources
erroneously calculated as undetectable will be omitted from the
analysis and, therefore, will not be included in the calculation of
the reliability ratio.  These catalog comparisons imply a reliability of $\gtrsim 99\%$ for sources detected above a detection SNR of 5.

\begin{figure}
\begin{center}
\includegraphics[width=0.9\linewidth]{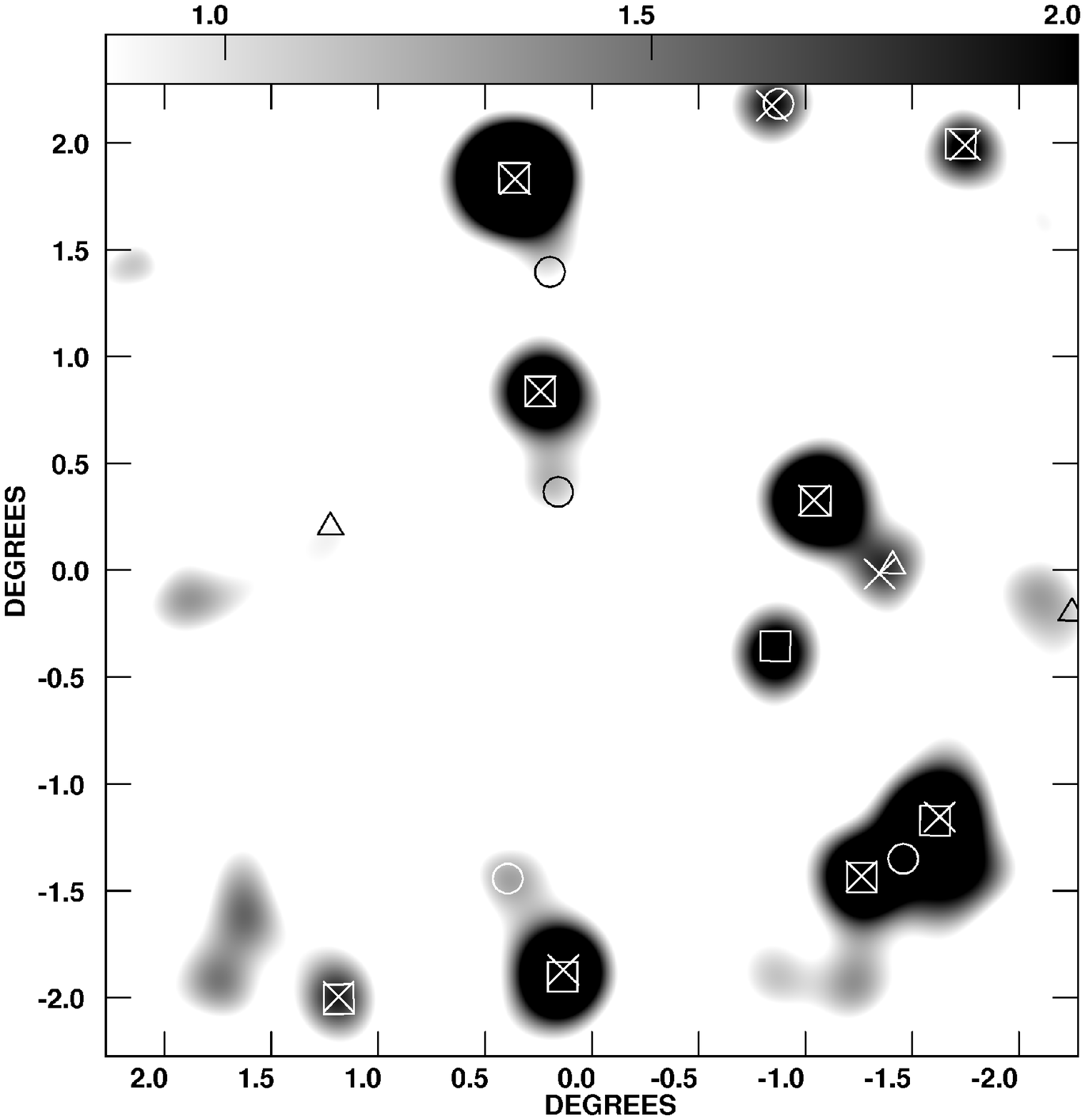}
\end{center}
\caption{Grey scale image of a TGSS field (field R33D18), with positions
of MWA and MRC sources overlaid.  Only pixels with ${\rm SNR} > 4$ are plotted,
and the mapping of pixels to grey scale is shown
by the scale in ${\rm Jy}~{\rm beam}^{-1}$ at the top of the image.  Positions of MWA sources with 
${\rm Detection~Threshold} > 5$ are plotted with a square, of MWA sources with $4 < {\rm Detection~Threshold} \leq 5$ are plotted with
a circle, of MWA source with $3 < {\rm Detection~Threshold} \leq 4$ with a triangle, and of MRC sources with an X.
All MWA and MRC sources in this field have a counterpart in the TGSS image.
There are five sources in the field that are detected in the MWA and TGSS surveys,
but not in the MRC.\label{fig:TGSS}}
\end{figure}

For the reliability comparison using the TGSS, we used the maps from TGSS
Data Release 2 available at the time of our analysis.  Although the baseline distribution of the GMRT
is substantially different from that of the MWA-32T, the GMRT has a compact central array consisting of fourteen antennas within an
area of radius 500 meters \citep{Swarup1991} that leads to substantial
overlap with the MWA-32T regarding the region of the $uv$ plane that
was sampled.  Twenty-seven TGSS fields overlapped the EoR2 field; none
overlapped the Hydra~A field.  We modeled the effects of source
blending by the large MWA beam by convolving the CLEANed and restored
TGSS maps to a Gaussian full width at half maximum resolution of
$12'$.  Care was taken in the convolution to preserve the flux
density scale. The resulting convolved images have typical RMS surface brightness fluctuations of $\sim$0.4~Jy/beam in regions free of sources.  Sources in
these maps were identified by taking all pixels above $4 \sigma$ and
associating a source with each island of bright pixels.  The resulting
TGSS source catalog was compared to those sources in the MWA list with
flux density greater than the $4 \sigma$ level in the TGSS field, and
thus expected to be detected in the TGSS field.  Pairs of sources in
the two catalogs coincident within $10'$ were recorded as
sources detected in both surveys.  MWA sources without a TGSS
counterpart were counted as non-detections.  Reliability values are
presented in Tables~\ref{tab:cum_reliability} and
\ref{tab:dif_reliability}.  These values are ratios for each MWA detection SNR
bin of the number of TGSS detections to the number of MWA sources
expected to be detected in the TGSS field, and are, of course, a
function of the threshold chosen in the TGSS maps.  As an example of the
TGSS comparison we present Figure~\ref{fig:TGSS}, which plots the
positions of MWA and MRC sources on a grey scale image of a convolved
TGSS field. It is important to note that these reliability estimates
solely test for the presence of a source coincident with the reported
position, and do not speak to the fidelity of the fluxes of these
sources or whether the MWA sources are due to single objects or blends
of multiple fainter objects.

\begin{deluxetable*}{ccccccc}
\tablecaption{Cumulative Source Reliability}
\tabletypesize{\scriptsize}
\tablewidth{0pt}
\tablehead{
\colhead{Detection SNR} & \multicolumn{2}{c}{MRC Reliability} & \multicolumn{2}{c}{VLSS Reliability} & \multicolumn{2}{c}{TGSS Reliability} \\
\colhead{Threshold} & \colhead{$N_{\rm detected}/N_{\rm detectable}$} & \colhead{R}  & \colhead{$N_{\rm detected}/N_{\rm detectable}$} & \colhead{R}  &\colhead{$N_{\rm detected}/N_{\rm detectable}$} & \colhead{R}\\
}
\startdata
$>3$ & 488/589   &  $82\%$           & 1215/1312 & 91\% & 183/197 & 93\%\\
$>4$ & 421/444    & $95\%$           & 826/839 & 98\%    & 132/133  & 99\%\\
$>5$ & 349/357     & $98\%$          & 575/579 & 99\%    & 85/85 & 100\%\\
$>7$ & 257/259    &  $99\%$          & 325/326 & 100\%  & 49/49 &  100\%\\
$>10$ & 167/167     & $100\%$      & 173/173 & 100\%  & 26/26 & 100\%\\
\enddata
\tablecomments{The reliability is assessed by comparing the MWA source list with the MRC catalog, VLSS catalog, and convolved TGSS maps as descriged in Section~\ref{sec:reliability}.  The MRC and VLSS comparisons are made by extrapolating the MWA source flux to 408~MHz and 74~MHz respectively, assuming a spectral index of $\alpha=-0.8$, to assess the detectability of the MWA source in the catalogs.  The TGSS results are based on using sources in convolved TGSS maps above $4 \sigma$ significance. The reliability percentages, $R$, have been corrected for  false positives.}
\label{tab:cum_reliability}
\end{deluxetable*}

\begin{deluxetable*}{ccccccc}
\tabletypesize{\scriptsize}
\tablecaption{Differential Source Reliability}
\tablewidth{0pt}
\tablehead{
\colhead{Detection SNR} & \multicolumn{2}{c}{MRC Diff. Reliability} & \multicolumn{2}{c}{VLSS Diff. Reliability} & \multicolumn{2}{c}{TGSS Diff. Reliability} \\
\colhead{Threshold (DT)} & \colhead{$N_{\rm detected}/N_{\rm detectable}$} & \colhead{R}  & \colhead{$N_{\rm detected}/N_{\rm detectable}$} & \colhead{R}  &\colhead{$N_{\rm detected}/N_{\rm detectable}$} & \colhead{R}\\
}
\startdata
$3<{\rm DT} \leq 4$ & 67/145 &    $44\%$ & 389/473 & 78\% & 51/64 & 80\%\\
$4<{\rm DT} \leq 5$ & 72/87  &  $82\%$    &  251/260 & 96\% & 47/48 & 98\%\\
$5<{\rm DT} \leq 6$ & 43/48   &  $89\%$   & 142/144 & 98\% & 20/20 & 100\%\\
$6<{\rm DT} \leq 7$ & 49/50  &   $98\%$   & 108/109 & 99\% & 16/16 & 100\% \\
$7<{\rm DT} \leq 8$ & 33/34  &   $97\%$   & 68/69 & 98\% & 15/15 & 100\%\\
$8<{\rm DT} \leq 9$ & 31/32  &   $97\%$   & 44/44 & 100\% & 7/7  & 100\%\\
$9<{\rm DT} \leq 10$ & 26/26  &  $100\%$ & 40/40 & 100\% & 1/1  & 100\%\\
\enddata
\tablecomments{The reliability is assessed by comparing the MWA source list with the MRC catalog, VLSS catalog, and convolved TGSS maps for various ranges of detection SNR threshold (DT) as discussed in Section~\ref{sec:reliability}.  The MRC and VLSS comparisons are made by extrapolating the MWA source flux to 408~MHz  and 74~MHz respectively, assuming a spectral index of $\alpha=-0.8$, to assess the detectability of the MWA source in the catalogs.  The TGSS results are based on using sources in convolved TGSS maps above $4 \sigma$ significance. The reliability percentages, $R$, have been corrected for false positives.}
\label{tab:dif_reliability}
\end{deluxetable*}

\subsection{Completeness of the Source Catalog\label{sec:completeness}}

We used the Culgoora source list to assess the completeness of the MWA
catalog presented in this paper.  We chose the Culgoora list because
it includes observations done at 160 MHz, a frequency not
far from the midpoint of the MWA band and because its synthesized beam
 is similar in size to that of the MWA-32T.  For each Culgoora source
within a field observed by the MWA we used the position-dependent RMS
noise in the MWA full-band averaged maps to evaluate its detectability in the MWA map.  Culgoora sources which should
be detectable above a specified SNR level in the MWA images were then
checked for a matching source within $10'$ in the MWA catalog.  Since we used the
Culgoora source list to assess the completeness, the results are only
valid down to a level comparable to the lowest Culgoora fluxes of
$\sim1.2$~Jy.  The completeness ratio was calculated similarly to the reliability described in Section~\ref{sec:reliability}:
\begin{equation}
C=\frac{N_{\rm detected} / N_{\rm detectable} -f}{1-f},
\end{equation}
where $C$ is the completeness percentage,  $N_{\rm detectable}$ is the number of Culgoora sources which we believe should have been detectable in the MWA source list , $N_{\rm detected}$ is the number of detectable Culgoora sources for which we found counterparts in the MWA list, and $f$ is the false source coincidence fraction calculated from the MWA catalog source density using the $10'$ source matching criteria.  The results are presented in Table~\ref{tab:completeness}.

The completeness was analyzed separately for sources within inner and
outer regions separated by a circle of radius $r = 18^\circ$
around the field center.  All Culgoora sources within the inner region
which did not have a corresponding detection in the MWA source list
were inspected and found to coincide with a local maximum in the map,
implying the completeness is limited by the robustness of the source
extraction algorithm and the flux calibration rather than the
intrinsic map quality.  It is important to note that because the MWA
maps have a sensitivity that varies strongly across the field, this
completeness value does not specify a flux limit to the catalog, but
rather assesses the efficacy of the source extraction.  As with the above reliability estimate, variability and incompleteness act to make this a lower limit on the true completeness.

\begin{deluxetable*}{ccccc}
\tablecaption{Source List Completeness}
\tablewidth{0pt}
\tablecolumns{5}
\tablehead{
\colhead{Field} & \multicolumn{2}{c}{$r<18^\circ$} & \multicolumn{2}{c}{$r>18^\circ$}\\
\colhead{Name} &$N_{\rm detected}/N_{\rm detectable}$ & Completeness &$N_{\rm detected}/N_{\rm detectable}$& Completeness \\
}
\startdata
\multicolumn{5}{l}{\em Detection SNR Level $\geq5$ }\\
Hydra~A & 56/63 &    $89\%$ &  36/44 &    $82\%$ \\
EoR2 & 72/77   &  $93\%$  & 49/58  &  $84\%$\\
\\
\multicolumn{5}{l}{\em Detection SNR Level $\geq3$ }\\
Hydra~A & 62/63 & 98\% & 52/67 & 77\% \\
EoR2 & 75/77 & 97\% & 61/66 & 92\% \\
\enddata
\tablecomments{The completeness as assessed by a comparison with the Culgoora source list \citep{Slee1995}, as described in Section~\ref{sec:completeness}.  The minimum source flux in the Culgoora list is $\sim$1.2~Jy, so these results are only valid for sources brighter than this level.  We view these completeness estimates as a lower limit on the catalog completeness --  source variability or flux errors in the Culgoora measurement will decrease the calculated completeness ratio.  Due to the varying sensitivity across the MWA field, the completeness is calculated relative to the local noise in the MWA source detection map, rather than an absolute flux level. The completeness percentages have been corrected for false positives as described in Section~\ref{sec:completeness}.  Analyzing the source counts in the field (see Section~\ref{sec:source_counts}) indicates that the source list is complete above $\sim$2~Jy.}
\label{tab:completeness}
\end{deluxetable*}

\subsection{Source Counts and Correlation Function\label{sec:source_counts}}

\begin{figure*}
\includegraphics[width=0.48\linewidth]{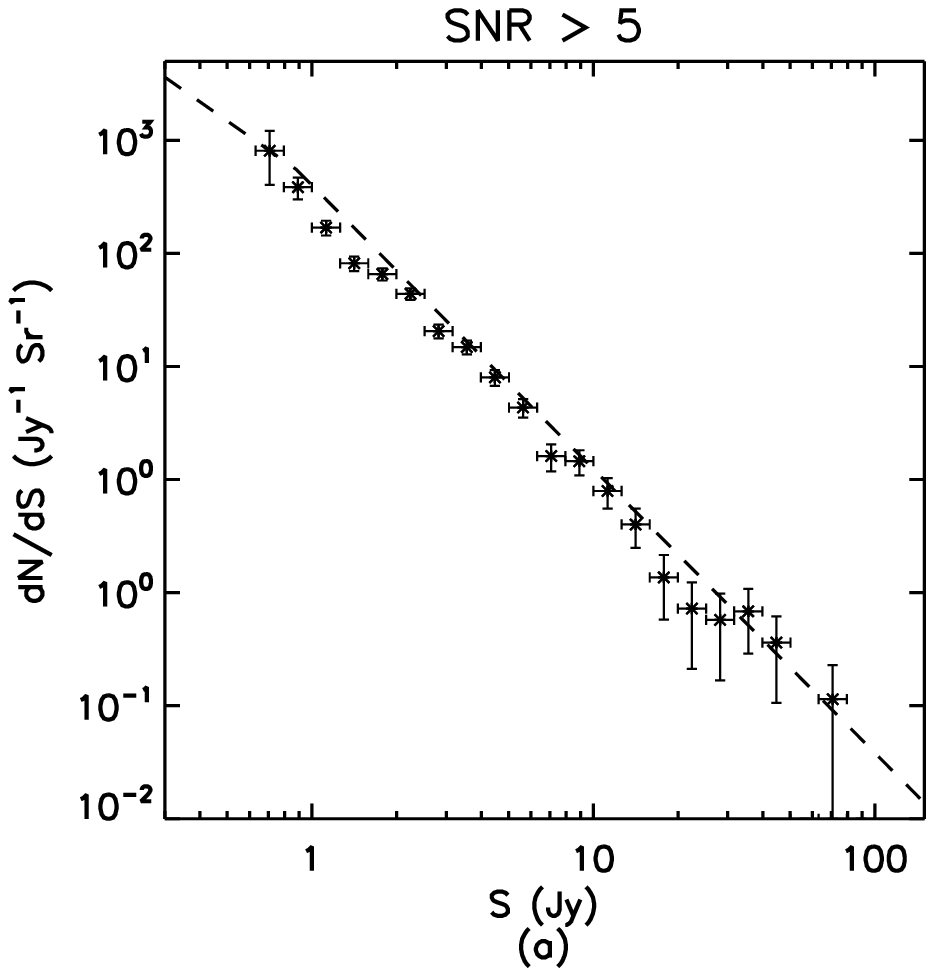}
\includegraphics[width=0.48\linewidth]{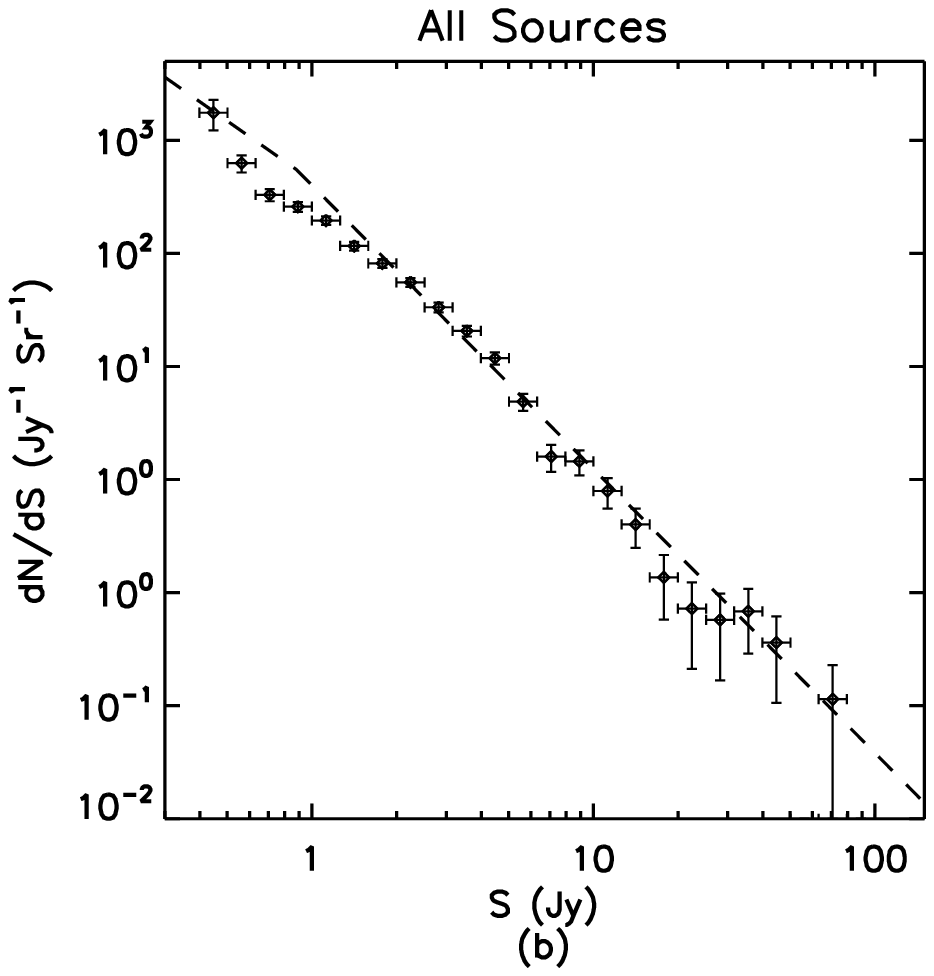}
\caption{Differential source counts histograms from the MWA EoR2 field, calculated using both the high reliability catalog (a) and the full list of source candidates (b).  The noise maps were used to correct for the effective area surveyed in each bin.  Poisson error bars are assigned based on the number of counts in each bin.  No Eddington bias correction is applied.  The source counts model from \citet{DiMatteo2002} is shown for reference with a dashed line.  We believe the deviations from the \citet{DiMatteo2002} model below $S\sim3~{\rm Jy}$ are not due to an intrinsic change in the source counts distribution at this scale, but are instead due to incompleteness of the source catalog in this flux range. \label{fig:NS}}
\end{figure*}

Radio source counts provide another useful diagnostic
test to assess the quality of the catalog and consistency with
previous works.  As discussed in Section~\ref{sec:results}, \citet{DiMatteo2002} fit the 151~MHz 6C survey
results of \citet{Hales1988} to obtain a power law model for
radio source counts.   The fit is valid up to $\sim10$ Jy, but the actual 6C counts
fall somewhat below the fit at the high end of the range.  Using the MWA-32T
catalog generated from the EoR2 field, we calculate the differential source counts, using the noise map of the field to correct for the
effects of sensitivity variations on the effective survey area for
different flux values (i.e., bright sources can be detected over a
larger area than faint sources).  No Eddington bias correction is applied (to correct for the artificial enhancement of faint sources due to noise in the map) and the error bars are calculated from the square root of the number of counts in each bin.  These source counts are shown in Figure~\ref{fig:NS}, along with the expected source counts from integrating the \citet{DiMatteo2002} model.  We note that more sophisticated models of 150~MHz source counts models have been described by, e.g., \citet{Wilman2008} and \citet{Jackson2005}.  These models have different behavior at the sub-Jansky flux levels that have been probed by high resolution, deep, narrower field-of-view studies such as those described in \citet{Intema2011} and \citet{IshwaraChandra2010}.  We compare our results with the \citet{DiMatteo2002} source counts model because it is commonly used as a basis for studies of EoR foregrounds and sensitivities.

The results from the catalog presented in this work agree with the \citet{DiMatteo2002} model above flux levels of $\sim$2~Jy.  Below this level, the MWA-32T source counts diverge from the model, likely because of the incompleteness of the MWA source extraction for low flux sources.  A power law fit to the EoR2 field source counts above 2~Jy yields $dn/dS= (3500 \pm 500) (S/1~{\rm Jy})^{-2.59 \pm 0.09}~{\rm sources}~{\rm Jy}^{-1}~{\rm sr}^{-1}$ for sources with a detection SNR greater than 5.  Fitting for a power law to the full list of sources in the field down to a detection SNR of 3 yields $dn/dS= (5700 \pm 700) (S/1~{\rm Jy})^{-2.76 \pm 0.08}~{\rm sources}~{\rm Jy}^{-1}~{\rm sr}^{-1}$.

\begin{figure} 
\includegraphics[width=0.75\linewidth,angle=270]{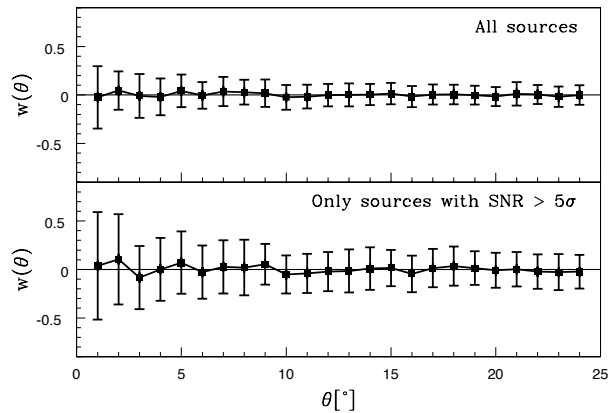}
\caption{Measured angular correlation function, $w(\theta)$, 
	 of the full MWA catalog. The angular correlations 
	 are calculated using sources with fluxes above $S\sim3~{\rm Jy}$. 
        The top panel shows $w(\theta)$ calculated from all sources 
        in our final catalog, while the bottom panel shows 
        $w(\theta)$ calculated only from sources detected 
        above a detection SNR threshold of 5. As expected, the results are consistent 
        with zero correlation. 
       \label{fig:wtheta}} 
\end{figure}

As an additional test for systematic effects, we have constructed the angular two-point correlation function, $w(\theta)$, of the sources in our catalog.  This correlation function can show systematic effects that manifest themselves on characteristic angular scales in the catalog -- see, e.g. \citet{Blake2002}.  We measure $w(\theta)$ using the estimator defined by  \citet{Hamilton1993}:
\begin{equation}
w(\theta) = \frac{DD(\theta) RR(\theta)}{DR(\theta)} - 1,
\end{equation}
where $DD(\theta)$ is the measured angular autocorrelation function from the MWA source catalog, $RR(\theta)$ is the autocorrelation function calculated using a simulated ``mock'' catalog, and $DR(\theta)$ is the cross-correlation between the MWA and the mock catalog.  We generate an ensemble of 100 mock catalogs and evaluate the correlation function with each one separately in order to produce a set of normally distributed estimates of $w(\theta)$.  Each mock catalog is produced using an 
approach developed to simulate point sources at  CMB 
and FIR frequencies \citep{Argueso2003,GonzalezNuevo2005}, but tailored 
specifically for the MWA experiment \citep{DeOliveiraCosta2012}, i.e., we drew sources from the observed MWA-32T source counts distribution described above in accordance with the expected low-frequency source clustering statistics \citep{DeOliveiraCosta2010,DeOliveiraCosta2010b}.   On the angular scales probed by this survey, no observable clustering is expected.  By constructing the mock catalogs in this manner, and correlating them with the observed distribution, the resulting estimate of $w(\theta)$ identifies any unexpected correlation which may be due to systematic errors in our survey or in our catalog construction procedure. 
Figure~\ref{fig:wtheta} shows our measurement of $w(\theta)$ 
above a flux limit of $S \approx 3~{\rm Jy}$ (black squares). 
Distances between the observed and/or simulated sources are measured 
in bins of  $\sim 1^\circ$, which is substantially above the MWA resolution.  The mean value of $w(\theta)$ is shown, along with uncertainties derived from calculating the covariance between $w(\theta)$ bins in the mock catalogs. As expected,  $w(\theta)$ is consistent with zero,  implying that there is no excess correlation in our catalog.

\subsection{Comparison with PAPER Results}

Comparing the present results from the MWA with results from PAPER
\citep{Parsons2010} is a particularly useful exercise, as both arrays
are new, broadband, wide field-of-view instruments with similar $uv$
coverage, and both are intended to be used to make EoR power spectrum
measurements.  We find that 43 of the sources found in the survey
(described in Section~\ref{sec:surveys}) of \citet{Jacobs2011} are
located in our survey region.  A search in our MWA-32T catalog reveals
unique counterparts within $30'$ (the PAPER beam size used by
\citealt{Jacobs2011} for source association) of the PAPER locations
for 31 out of these 43 sources, multiple counterparts in 11 cases, and
no counterpart in one case.

The one source with no MWA counterpart is $24^\circ$ from the center
of the MWA Hydra~A field, and corresponds to a local maximum in the
MWA image; however it was not detected by the automatic source finding
algorithm.  For each of the 10 sources with multiple MWA counterparts, an
estimate of the blended flux was obtained by summing the flux of all
MWA sources within the PAPER beam.  Other than Hydra~A, all 41 PAPER
fluxes are consistent with the MWA
blended fluxes (the Hydra~A flux reported in the PAPER catalog was
corrupted by the filtering used in the PAPER analysis,
\citealt{Jacobs2011pc}, private communication).  A weighted average of
the ratios of the MWA and PAPER source fluxes yields the average ratio
$<S_{\rm MWA}/S_{\rm PAPER}>\ = 1.17 \pm 0.10$.  However, we note that the PAPER flux scale was set using measurements of two calibration sources from the Culgoora source list, whereas the MWA-32T flux scale was set using a fit to an ensemble of Culgoora and MRC measurements.  \citet{Slee1977} note that the Culgoora flux scale may be depressed by $10\%$, with additional flux uncertainties of between 13\% and 39\% for individual source measurements.  If these Culgoora flux uncertainties are taken into account as potential errors on the PAPER flux scale, then the significance of the difference between the MWA and PAPER flux scales is decreased.

The standard deviation of the MWA to PAPER flux ratios after correcting for the different flux scales is $\sim$25\%.  This
is smaller than our estimate of the MWA flux uncertainties based on flux predictions from the MRC and Culgoora measurements.
This indicates that the flux comparison with the MRC and Culgoora lists may be
affected to a considerable extent by radio source variability or other systematic effects.

\begin{figure}
\plotone{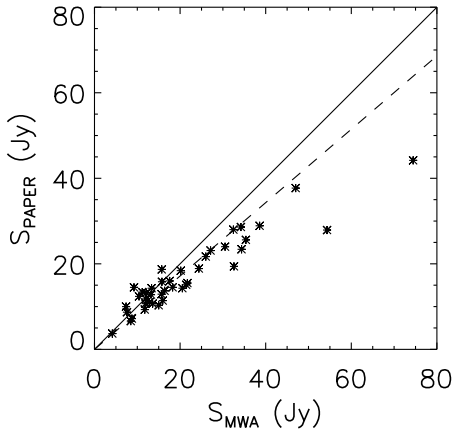}
\caption{Comparison of the fluxes presented in this work with those from the 145~MHz PAPER source list presented in \citet{Jacobs2011}.  Error bars have been omitted from the plot for clarity.  A total of 43 sources from the PAPER list are within the MWA field.  MWA sources, which are within $30'$ of a PAPER source are matched.  A total of 31 of the PAPER sources had unique counterparts, while 11 PAPER sources matched with multiple MWA sources and 1 PAPER source did not have a detected MWA counterpart (although there is a local maximum in the MWA map at the location of the PAPER source).  The MWA fluxes are calculated by summing the flux of all MWA sources which match with a PAPER source, and scaling the flux to a frequency of 145~MHz assuming a spectral index of $\alpha=-0.8$ ($S\propto \nu^{-\alpha}$).  The solid line shows the unity flux-ratio locus; the MWA sources are on average 17\% brighter than the PAPER sources.  This fitted flux ratio locus is plotted as a dashed line.  \label{fig:paperfluxes}}
\end{figure}

\subsection{Candidate Ultra Steep Spectrum Sources}

\begin{figure}
\plotone{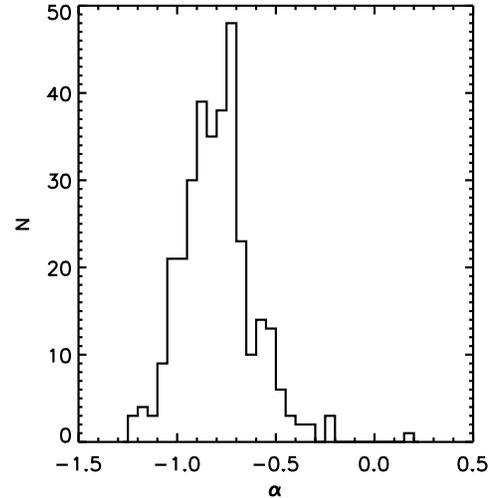}
\caption{Distribution of spectral indices ($S_{\nu}\propto \nu^\alpha$) between 154.25~MHz and 4.85~GHz for sources identified in this paper, based on a comparison with the Parkes-MIT-NRAO catalog \citep{Griffith1993}.  To avoid issues of source confusion and blending, only sources which could be unambiguously associated with single PMN counterparts are included in the histogram.\label{fig:specindex}}
\end{figure}

``Ultra Steep Spectrum'' (USS) radio sources form a compelling class of candidate high redshift radio sources \citep{DeBreuck2000, DiMatteo2004, DeBreuck2002, Broderick2007}.  Low frequency radio observations are particularly sensitive to these objects (see, e.g. \citealt{Pedani2003}).  We have conducted an analysis of the sources detected in the MWA fields in an attempt to identify additional USS radio sources.  A table of candidates is presented in Table~\ref{tab:uss}.  We calculated spectral indices by using the PMN 4.85~GHz survey \citep{Griffith1993} together with the MWA full-band average flux measurements.  We associated MWA sources with PMN counterparts if their positions were coincident within $5'$.  To avoid source confusion and blending issues, we excluded any MWA sources with more than one PMN counterpart within a $30'$ radius.  A total of 331 sources were identified for which we could unambiguously identify a counterpart and extract a spectral index.  A histogram of the spectral indices are shown in Figure~\ref{fig:specindex}.  This histogram appears consistent with the low-frequency selected spectral index distributions obtained by \citet{DeBreuck2000}, and plotted in their Figure~7 (however, the \citealt{DeBreuck2000} analysis used slightly different frequencies).  Using a low-frequency selected distribution results in a sample which is significantly more sensitive to the USS sources.  

\begin{deluxetable*}{ccccccccc}
\tablecaption{Ultra-Steep Spectrum Source Candidates}

\tablehead{
\colhead{Name} & \colhead{RA} & \colhead{DEC} & \colhead{$S_{\rm MWA,avg}$}& \colhead{$S_{\rm PMN}$} & \colhead{$\alpha_{\rm PMN}$} & \colhead{$S_{\rm MRC}$} & \colhead{$\alpha_{\rm MRC}$} & \colhead{Detection}\\
&&&&&&&&\colhead{SNR Level} \\
}

\tabletypesize{\scriptsize}

\startdata
J1009$-$1207 & 10h09m19s & $-12^\circ07'46''$ & 10.69$\pm$3.21 & 0.17$\pm$0.01 & -1.20$\pm$0.09 & 3.23$\pm$0.11 & -1.23$\pm$0.31 & 52.5 \\
J1032$-$3421 & 10h32m60s & $-34^\circ21'19''$ & 17.38$\pm$5.31 & 0.27$\pm$0.02 & -1.21$\pm$0.09 & 5.59$\pm$0.25 & -1.17$\pm$0.32 & 15.5 \\
J1042$+$1201 & 10h42m56s & $+12^\circ01'30''$ & 15.55$\pm$4.80 & \nodata & $<$-1.66$\pm$0.09 & 8.90$\pm$0.37 & -0.57$\pm$0.32 & 15.1\\
J1034$+$1111 & 10h34m13s & $+11^\circ11'23''$ & 5.89$\pm$1.92 & \nodata & $<$-1.38$\pm$0.09 & 3.80$\pm$0.16 & -0.45$\pm$0.34 & 8.3 \\
J1000$+$1400 & 10h00m16s & $+14^\circ00'17''$ & 7.54$\pm$2.51 & \nodata & $<$-1.45$\pm$0.10 & 3.06$\pm$0.10 & -0.93$\pm$0.34 & 6.3 \\
J0831$-$2922 & 08h31m24s & $-29^\circ22'26''$ & 3.65$\pm$1.24 & \nodata & $<$-1.24$\pm$0.10 & 1.51$\pm$0.06 & -0.91$\pm$0.35 & 5.7 \\
J1007$+$1246 & 10h07m28s & $+12^\circ46'50''$ & 7.46$\pm$2.52 & \nodata & $<$-1.45$\pm$0.10 & \nodata & \nodata & 5.6 \\
J0855$+$0552 & 08h55m18s & $+05^\circ52'50''$ & 4.11$\pm$1.40 & 0.06$\pm$0.01 & -1.23$\pm$0.12 & 1.62$\pm$0.07 & -0.96$\pm$0.35 & 5.4 \\
J0834$-$3443 & 08h34m29s & $-34^\circ43'22''$ & 6.14$\pm$2.12 & \nodata & $<$-1.40$\pm$0.10 & \nodata & \nodata & 5.0 \\
\hline
J0828$-$3201 & 08h28m14s & $-32^\circ01'26''$ & 4.68$\pm$1.66 & \nodata & $<$-1.32$\pm$0.10 & \nodata & \nodata & 4.8 \\
J1001$+$1108 & 10h01m01s & $+11^\circ08'19''$ & 3.98$\pm$1.37 & \nodata & $<$-1.27$\pm$0.10 & 1.57$\pm$0.06 & -0.96$\pm$0.36 & 4.6 \\
J0832$-$3326 & 08h32m38s & $-33^\circ26'00''$ & 5.14$\pm$1.84 & \nodata & $<$-1.34$\pm$0.10 & 2.30$\pm$0.09 & -0.83$\pm$0.37 & 4.4 \\
J1008$+$1201 & 10h08m11s & $+12^\circ01'07''$ & 4.48$\pm$1.65 & \nodata & $<$-1.30$\pm$0.11 & \nodata & \nodata & 4.2 \\
J1028$+$1158 & 10h28m32s & $+11^\circ58'58''$ & 3.56$\pm$1.30 & \nodata & $<$-1.24$\pm$0.11 & \nodata & \nodata & 4.1 \\
J1112$+$1112 & 11h12m45s & $+11^\circ12'45''$ & 4.88$\pm$1.92 & \nodata & $<$-1.33$\pm$0.11 & 1.45$\pm$0.07 & -1.25$\pm$0.41 & 3.8 \\
J1021$+$1303 & 10h21m34s & $+13^\circ03'55''$ & 3.76$\pm$1.47 & \nodata & $<$-1.25$\pm$0.11 & \nodata & \nodata & 3.7 \\
J1034$+$1428 & 10h34m18s & $+14^\circ28'48''$ & 4.68$\pm$1.85 & \nodata & $<$-1.32$\pm$0.11 & \nodata & \nodata & 3.5 \\
J1104$+$1103 & 11h04m21s & $+11^\circ03'31''$ & 5.35$\pm$2.01 & \nodata & $<$-1.36$\pm$0.11 & \nodata & \nodata & 3.4 \\
J0827$-$3322 & 08h27m30s & $-33^\circ22'43''$ & 3.61$\pm$1.47 & \nodata & $<$-1.24$\pm$0.12 & \nodata & \nodata & 3.4 \\
J0918$+$1226 & 09h18m48s & $+12^\circ26'31''$ & 5.67$\pm$2.34 & \nodata & $<$-1.37$\pm$0.12 & 2.02$\pm$0.09 & -1.06$\pm$0.43 & 3.3 \\
J1027$+$1347 & 10h27m10s & $+13^\circ47'25''$ & 3.94$\pm$1.52 & \nodata & $<$-1.27$\pm$0.11 & \nodata & \nodata & 3.3 \\
J1114$+$1048 & 11h14m39s & $+10^\circ48'59''$ & 4.28$\pm$1.71 & \nodata & $<$-1.29$\pm$0.12 & \nodata & \nodata & 3.2 \\
J0843$+$1115 & 08h43m31s & $+11^\circ15'20''$ & 6.34$\pm$2.43 & \nodata & $<$-1.40$\pm$0.11 & 1.34$\pm$0.07 & -1.60$\pm$0.40 & 3.2 \\
J1015$+$1141 & 10h15m39s & $+11^\circ41'05''$ & 3.16$\pm$1.31 & \nodata & $<$-1.20$\pm$0.12 & \nodata & \nodata & 3.2 \\
J0929$+$1133 & 09h29m20s & $+11^\circ33'39''$ & 5.71$\pm$2.13 & \nodata & $<$-1.37$\pm$0.11 & 1.69$\pm$0.08 & -1.25$\pm$0.39 & 3.1 \\
J1019$+$1405 & 10h19m53s & $+14^\circ05'41''$ & 3.69$\pm$1.46 & \nodata & $<$-1.25$\pm$0.11 & 0.87$\pm$0.06 & -1.49$\pm$0.41 & 3.1 \\
J1026$+$1431 & 10h26m53s & $+14^\circ31'09''$ & 3.96$\pm$1.67 & \nodata & $<$-1.27$\pm$0.12 & \nodata & \nodata & 3.0 \\
J0918$+$1114 & 09h18m34s & $+11^\circ14'45''$ & 5.40$\pm$2.24 & \nodata & $<$-1.36$\pm$0.12 & \nodata & \nodata & 3.0 \\

\enddata
\tablecomments{This USS sample is created by matching sources in the MWA-32T catalog with uniquely corresponding sources in the 4.85~GHz PMN catalog.  Sources which have spectral indices of $\alpha <-1.2$ ($S \propto \nu^\alpha$) are identified as ultra-steep candidates.  The sources are sorted in order of decreasing detection SNR level.  MWA sources which are within $5'$ of a PMN source, and have no other PMN counterparts within $30'$ are matched as counterparts and used to calculate the spectral index.  Additionally, for MWA sources with no PMN source within $1^\circ$, a spectral index limit is calculated using the PMN flux limit of 50~mJy.    For comparison, the ultra-steep candidates are matched with MRC candidates within $15'$ (sources with multiple matches within $30'$ are excluded), and an additional spectral index is calculated.}
\label{tab:uss}
\end{deluxetable*}

We choose a spectral index cutoff of $\alpha \le -1.2$ for USS source candidates, and find 3 sources which match the criteria.  All three sources have counterparts in the MRC, and the source MWA J1032-3421 has a counterpart in both the PAPER and Culgoora source lists (although the other two USS candidates do not).  A further 33 sources are identified which have no counterpart in the PMN catalog within $1^\circ$.  Using the PMN catalog limiting flux of 50~mJy \citep{Griffith1993} for these sources, we find that 25 of these 33 sources have an inferred spectral index of $\alpha \le -1.2$.  Of these 25 sources, 11 have unique counterparts in the MRC (one additional source has multiple counterparts), 9 have counterparts in the Culgoora source list, and none have counterparts in the PAPER source list.

A comparison between the USS source list of \citet{DeBreuck2000} and the source list identified in this work finds only 1 source from their list which matches a MWA USS candidate within $15'$: MWA J1032-3421.  They select this source from an analysis of the MRC and PMN samples, and find a spectral index of $-1.23\pm0.04$, which is consistent with the MWA-32T measurements.   There are an additional 21 sources from the \citet{DeBreuck2000} sample which match MWA sources, however only three of these MWA sources matched uniquely with a PMN counterpart: MWA J1133-2717, MWA J0941-1627, and MWA J0937-2243.  These three sources were identified as USS sources in the ``TN'' sample of \citet{DeBreuck2000}, which used measurements at 365~MHz and 1.4~GHz, but they did not meet the USS criteria in the MWA-PMN comparison.  The flatter MWA-PMN spectra may indicate a high-frequency turnover in the source spectrum, similar to that noted for J0008-421 in \citet{Jacobs2011}.  This interpretation is supported by 74~MHz VLSS measurements of these sources \citep{Cohen2007}, which imply a spectral flattening at low frequencies, although further follow-up will be important to definitively establish this behavior.

It is important to note that these MWA sources are only candidates, and should not be treated as definitive ultra-steep spectrum sources.  Flux calibration and measurement errors as well as blending issues due to the low MWA resolution and time variability may result in a reclassification of these sources upon more detailed investigation.  Additionally any time variability or errors introduced by the different catalog resolutions and fitting algorithms will add to errors in this candidate list.  The MWA instrument is under continued development, and the fidelity of these studies will improve as the systematics are better understood.  However, this candidate list serves as a good basis for more detailed follow-up and investigation.

%

\section{Conclusions and Future Work}

The goals of this work were to verify the performance of the MWA subsystems and the MWA-32T system, to explore techniques for future EoR experiments, and to deepen our understanding of the radio sky at these frequencies.  The analysis and results presented in this paper served to help commission MWA-32T and represent an assessment of its performance.  Specifically, our ability to successfully solve for antenna gains in spite of the wide field of view and direction-dependent primary beam increases our confidence in our ability to calibrate the full 128-tile MWA array.  The high level of agreement  of our position measurements with existing source position measurements provides further verification of our understanding of the geometry of the array and the calibration procedures as well as our expectations for ionospheric effects on scales relevant to the MWA-32T array.  In short, the high fidelity, wide-field images produced in this analysis helps to build confidence that the MWA will be able to achieve its design goals.  The measured fluxes in the maps agree with expectations from previous catalogs with a scatter of about 30\%.  The magnitude of these flux residuals is similar to what is reported from other pathfinder low-frequency arrays, however it is significantly larger than expected based on the noise in the maps.  Further work is needed to understand the effects that cause these flux discrepancies.  

Future EoR experiments with the MWA will require long integrations, which will require the combination of data from many pointings.  We have shown in this work that these data can be corrected for primary beam effects and weighted averages can be formed to increase the sensitivity and fidelity of the resulting maps.  In this initial exploration we made certain simplifying assumptions that allowed us to do this analysis in the image domain.  Future, deeper, investigations will require more sophisticated techniques that account for differences between individual antenna elements.  Developing and verifying these techniques should also be a priority.

A distinguishing feature of the images presented in this paper is the degree to which they are confusion-limited.  The effects of source confusion are not expected to be an issue for EoR experiments.  Simulations indicate, though it remains to be shown in practice, that subtracting the brightest sources, and treating the fainter sources as smooth contributors to each pixel will allow an adequate separation of foregrounds from the EoR signal (see, e.g. \citealt{Liu2011}).  For other science goals, however, it is quite clear that for the low-frequency arrays planned for the near future, the confusion limit of continuum images will be reached in relatively short integration times.  Thus, much of the science will have to contend with the difficulties of measurements in crowded fields.  We have presented the results of an automatic source extraction algorithm.  While the results are quite good, with reliability over 90\%, further development of algorithms that extract more accurate source models from crowded fields (such as the algorithm presented in \citealt{Hancock2012}) should be a priority.  

The source list presented in this paper serves as the deepest catalog of radio sources generated from a blind search in this region of the southern sky in this frequency band and is the first MWA characterization of the EoR point source foreground in the MWA EoR field.  Further refinement of statistical descriptions of this component of the foreground should be possible, and represent a promising avenue for future work.  The production of this large, high quality survey with only $\sim$25~hours of data highlights the survey power of this instrument.  The techniques used to survey these two fields can be extended to complete an MWA all-sky survey.  It is important to note that because this survey was carried out near the source confusion limit, many of the fainter sources in the sample are likely blends of multiple sources.  The USS candidates identified in this work, although likely affected by this blending, serve as a set of candidates for high-redshift radio sources, and are good targets for follow-up with higher resolution low-frequency instruments.  

The MWA is in the process of a buildout to a 128-tile (128T) interferometer.  The 128T array will have four times the number of collecting elements as the 32T array, and the maximum baseline length will be increased to $\sim$3~km.  The higher resolution of the 128T array will yield maps with a lower source confusion limit, and will enable a deeper survey of this sky field with a reduced number of blended sources.  With the added sensitivity, the 128T array will still be able to rapidly produce source confusion limited maps.  The techniques developed as part of the work described herein will allow evaluation of the 128T instrument as it is commissioned.  As the quality of the sky model at these frequencies improves, the increased calibration accuracy and ability to subtract foregrounds from the data will move us closer towards the goal of detecting the 21-cm signal during reionization. 

\acknowledgements
This scientific work uses data obtained from the Murchison Radio-astronomy Observatory.  We acknowledge the Wajarri Yamatji people as the traditional owners of the Observatory site.  Support for this work comes from the Australian Research Council (grant numbers LE0775621 and LE0882938), the National Science Foundation (grant numbers AST-0457585, AST-0821321, AST-0908884, AST-1008353 and PHY-0835713), the U.S. Air Force Office of Scientific Research (grant number FA9550-0510247), the Australian National Collaborative Research Infrastructure Strategy, the Australia India Strategic Research Fund, the Smithsonian Astrophysical Observatory, the MIT School of Science, MIT's Marble Astrophysics Fund, the Raman Research Institute, the Australian National University, the iVEC Petabyte Data Store, the NVIDIA sponsored CUDA Center for Excellence at Harvard University, and the International Centre for Radio Astronomy Research, a Joint Venture of Curtin University of Technology and The University of Western Australia, funded by the Western Australian State government.  The Centre for All-sky Astrophysics is an Australian Research Council Centre of Excellence, funded by grant CE11E0001020.  The MRO is managed by the CSIRO, who also provide operational support to the MWA.  This research work has used the TIFR GMRT Sky Survey (http://tgss.ncra.tifr.res.in) data products.  CLW would like to thank Adrian Liu, and Leo Stein for helpful discussions and comments, as well as the Halleen family for their hospitality at Boolardy Station while working at the MRO.

\bibliography{21CM}

\begin{thebibliography}{85}
\expandafter\ifx\csname natexlab\endcsname\relax\def\natexlab#1{#1}\fi

\bibitem[{{Arg{\"u}eso} {et~al.}(2003){Arg{\"u}eso}, {Gonz{\'a}lez-Nuevo}, \&
  {Toffolatti}}]{Argueso2003}
{Arg{\"u}eso}, F., {Gonz{\'a}lez-Nuevo}, J., \& {Toffolatti}, L. 2003, \apj,
  598, 86

\bibitem[{{Baldwin} {et~al.}(1985){Baldwin}, {Boysen}, {Hales}, {Jennings},
  {Waggett}, {Warner}, \& {Wilson}}]{Baldwin1985}
{Baldwin}, J.~E., {Boysen}, R.~C., {Hales}, S.~E.~G., {Jennings}, J.~E.,
  {Waggett}, P.~C., {Warner}, P.~J., \& {Wilson}, D.~M.~A. 1985, \mnras, 217,
  717

\bibitem[{{Bernardi} {et~al.}(2009){Bernardi}, {de Bruyn}, {Brentjens},
  {Ciardi}, {Harker}, {Jeli{\'c}}, {Koopmans}, {Labropoulos}, {Offringa},
  {Pandey}, {Schaye}, {Thomas}, {Yatawatta}, \& {Zaroubi}}]{Bernardi2009}
{Bernardi}, G., {et~al.} 2009, \aap, 500, 965

\bibitem[{{Blake} \& {Wall}(2002)}]{Blake2002}
{Blake}, C., \& {Wall}, J. 2002, \mnras, 329, L37

\bibitem[{{Bock} {et~al.}(1999){Bock}, {Large}, \& {Sadler}}]{Bock1999}
{Bock}, D.~C.-J., {Large}, M.~I., \& {Sadler}, E.~M. 1999, \aj, 117, 1578

\bibitem[{{Bowman} {et~al.}(2009){Bowman}, {Morales}, \& {Hewitt}}]{Bowman2009}
{Bowman}, J.~D., {Morales}, M.~F., \& {Hewitt}, J.~N. 2009, \apj, 695, 183

\bibitem[{{Bowman} \& {Rogers}(2010)}]{Bowman2010}
{Bowman}, J.~D., \& {Rogers}, A.~E.~E. 2010, \nat, 468, 796

\bibitem[{{Broderick} {et~al.}(2007){Broderick}, {Bryant}, {Hunstead},
  {Sadler}, \& {Murphy}}]{Broderick2007}
{Broderick}, J.~W., {Bryant}, J.~J., {Hunstead}, R.~W., {Sadler}, E.~M., \&
  {Murphy}, T. 2007, \mnras, 381, 341

\bibitem[{{Chippendale}(2009)}]{Chippendale2009}
{Chippendale}, A.~P. 2009, PhD thesis, The University of Sydney

\bibitem[{{Cohen} {et~al.}(2007){Cohen}, {Lane}, {Cotton}, {Kassim}, {Lazio},
  {Perley}, {Condon}, \& {Erickson}}]{Cohen2007}
{Cohen}, A.~S., {Lane}, W.~M., {Cotton}, W.~D., {Kassim}, N.~E., {Lazio},
  T.~J.~W., {Perley}, R.~A., {Condon}, J.~J., \& {Erickson}, W.~C. 2007, \aj,
  134, 1245

\bibitem[{{Condon}(1974)}]{Condon1974}
{Condon}, J.~J. 1974, \apj, 188, 279

\bibitem[{{Condon}(1997)}]{Condon1997}
---. 1997, \pasp, 109, 166

\bibitem[{{Condon} {et~al.}(1998){Condon}, {Cotton}, {Greisen}, {Yin},
  {Perley}, {Taylor}, \& {Broderick}}]{Condon1998}
{Condon}, J.~J., {Cotton}, W.~D., {Greisen}, E.~W., {Yin}, Q.~F., {Perley},
  R.~A., {Taylor}, G.~B., \& {Broderick}, J.~J. 1998, \aj, 115, 1693

\bibitem[{{Condon} {et~al.}(1993){Condon}, {Griffith}, \&
  {Wright}}]{Condon1993}
{Condon}, J.~J., {Griffith}, M.~R., \& {Wright}, A.~E. 1993, \aj, 106, 1095

\bibitem[{{Cornwell} {et~al.}(2008){Cornwell}, {Golap}, \&
  {Bhatnagar}}]{Cornwell2008}
{Cornwell}, T.~J., {Golap}, K., \& {Bhatnagar}, S. 2008, IEEE Journal of
  Selected Topics in Signal Processing, 2, 647

\bibitem[{{Dagkesamanski{\u \i}} {et~al.}(2000){Dagkesamanski{\u \i}},
  {Samodurov}, \& {Lapaev}}]{Dagkesamanskii2000}
{Dagkesamanski{\u \i}}, R.~D., {Samodurov}, V.~A., \& {Lapaev}, K.~A. 2000,
  Astronomy Reports, 44, 18

\bibitem[{{Datta} {et~al.}(2010){Datta}, {Bowman}, \& {Carilli}}]{Datta2010}
{Datta}, A., {Bowman}, J.~D., \& {Carilli}, C.~L. 2010, \apj, 724, 526

\bibitem[{{De Breuck} {et~al.}(2000){De Breuck}, {van Breugel},
  {R{\"o}ttgering}, \& {Miley}}]{DeBreuck2000}
{De Breuck}, C., {van Breugel}, W., {R{\"o}ttgering}, H.~J.~A., \& {Miley}, G.
  2000, \aaps, 143, 303

\bibitem[{{De Breuck} {et~al.}(2002){De Breuck}, {van Breugel}, {Stanford},
  {R{\"o}ttgering}, {Miley}, \& {Stern}}]{DeBreuck2002}
{De Breuck}, C., {van Breugel}, W., {Stanford}, S.~A., {R{\"o}ttgering}, H.,
  {Miley}, G., \& {Stern}, D. 2002, \aj, 123, 637

\bibitem[{{de Oliveira-Costa} {et~al.}(in prep.){de Oliveira-Costa},
  {Bernardi}, \& {Gonzalez-Nuevo}}]{DeOliveiraCosta2012}
{de Oliveira-Costa}, A., {Bernardi}, G., \& {Gonzalez-Nuevo}, J. in prep.

\bibitem[{{de Oliveira-Costa} \& {Capodilupo}(2010)}]{DeOliveiraCosta2010}
{de Oliveira-Costa}, A., \& {Capodilupo}, J. 2010, \mnras, 404, 1962

\bibitem[{{de Oliveira-Costa} \& {Lazio}(2010)}]{DeOliveiraCosta2010b}
{de Oliveira-Costa}, A., \& {Lazio}, J. 2010, arXiv:1004.3167

\bibitem[{{de Oliveira-Costa} {et~al.}(2008){de Oliveira-Costa}, {Tegmark},
  {Gaensler}, {Jonas}, {Landecker}, \& {Reich}}]{DeOliveiraCosta2008}
{de Oliveira-Costa}, A., {Tegmark}, M., {Gaensler}, B.~M., {Jonas}, J.,
  {Landecker}, T.~L., \& {Reich}, P. 2008, \mnras, 388, 247

\bibitem[{{Di Matteo} {et~al.}(2004){Di Matteo}, {Ciardi}, \&
  {Miniati}}]{DiMatteo2004}
{Di Matteo}, T., {Ciardi}, B., \& {Miniati}, F. 2004, \mnras, 355, 1053

\bibitem[{{Di Matteo} {et~al.}(2002){Di Matteo}, {Perna}, {Abel}, \&
  {Rees}}]{DiMatteo2002}
{Di Matteo}, T., {Perna}, R., {Abel}, T., \& {Rees}, M.~J. 2002, \apj, 564, 576

\bibitem[{{Douglas} {et~al.}(1996){Douglas}, {Bash}, {Bozyan}, {Torrence}, \&
  {Wolfe}}]{Douglas1996}
{Douglas}, J.~N., {Bash}, F.~N., {Bozyan}, F.~A., {Torrence}, G.~W., \&
  {Wolfe}, C. 1996, \aj, 111, 1945

\bibitem[{{Furlanetto} \& {Briggs}(2004)}]{Furlanetto2004}
{Furlanetto}, S.~R., \& {Briggs}, F.~H. 2004, \nar, 48, 1039

\bibitem[{{Furlanetto} {et~al.}(2006){Furlanetto}, {Oh}, \&
  {Briggs}}]{Furlanetto2006}
{Furlanetto}, S.~R., {Oh}, S.~P., \& {Briggs}, F.~H. 2006, \physrep, 433, 181

\bibitem[{{Gonz{\'a}lez-Nuevo} {et~al.}(2005){Gonz{\'a}lez-Nuevo},
  {Toffolatti}, \& {Arg{\"u}eso}}]{GonzalezNuevo2005}
{Gonz{\'a}lez-Nuevo}, J., {Toffolatti}, L., \& {Arg{\"u}eso}, F. 2005, \apj,
  621, 1

\bibitem[{{Griffith} \& {Wright}(1993)}]{Griffith1993}
{Griffith}, M.~R., \& {Wright}, A.~E. 1993, \aj, 105, 1666

\bibitem[{{Griffith} {et~al.}(1994){Griffith}, {Wright}, {Burke}, \&
  {Ekers}}]{Griffith1994}
{Griffith}, M.~R., {Wright}, A.~E., {Burke}, B.~F., \& {Ekers}, R.~D. 1994,
  \apjs, 90, 179

\bibitem[{{Griffith} {et~al.}(1995){Griffith}, {Wright}, {Burke}, \&
  {Ekers}}]{Griffith1995}
---. 1995, \apjs, 97, 347

\bibitem[{{Hales} {et~al.}(1988){Hales}, {Baldwin}, \& {Warner}}]{Hales1988}
{Hales}, S.~E.~G., {Baldwin}, J.~E., \& {Warner}, P.~J. 1988, \mnras, 234, 919

\bibitem[{{Hales} {et~al.}(1993{\natexlab{a}}){Hales}, {Baldwin}, \&
  {Warner}}]{Hales1993b}
---. 1993{\natexlab{a}}, \mnras, 263, 25

\bibitem[{{Hales} {et~al.}(1990){Hales}, {Masson}, {Warner}, \&
  {Baldwin}}]{Hales1990}
{Hales}, S.~E.~G., {Masson}, C.~R., {Warner}, P.~J., \& {Baldwin}, J.~E. 1990,
  \mnras, 246, 256

\bibitem[{{Hales} {et~al.}(1993{\natexlab{b}}){Hales}, {Masson}, {Warner},
  {Baldwin}, \& {Green}}]{Hales1993a}
{Hales}, S.~E.~G., {Masson}, C.~R., {Warner}, P.~J., {Baldwin}, J.~E., \&
  {Green}, D.~A. 1993{\natexlab{b}}, \mnras, 262, 1057

\bibitem[{{Hales} {et~al.}(1991){Hales}, {Mayer}, {Warner}, \&
  {Baldwin}}]{Hales1991}
{Hales}, S.~E.~G., {Mayer}, C.~J., {Warner}, P.~J., \& {Baldwin}, J.~E. 1991,
  \mnras, 251, 46

\bibitem[{{Hales} {et~al.}(2007){Hales}, {Riley}, {Waldram}, {Warner}, \&
  {Baldwin}}]{Hales2007}
{Hales}, S.~E.~G., {Riley}, J.~M., {Waldram}, E.~M., {Warner}, P.~J., \&
  {Baldwin}, J.~E. 2007, \mnras, 382, 1639

\bibitem[{{Hamaker} {et~al.}(1996){Hamaker}, {Bregman}, \&
  {Sault}}]{Hamaker1996}
{Hamaker}, J.~P., {Bregman}, J.~D., \& {Sault}, R.~J. 1996, \aaps, 117, 137

\bibitem[{{Hamilton}(1993)}]{Hamilton1993}
{Hamilton}, A.~J.~S. 1993, \apj, 417, 19

\bibitem[{{Hancock} {et~al.}(2012){Hancock}, {Murphy}, {Gaensler}, {Hopkins},
  \& {Curran}}]{Hancock2012}
{Hancock}, P.~J., {Murphy}, T., {Gaensler}, B.~M., {Hopkins}, A., \& {Curran},
  J.~R. 2012, arXiv:1202.4500

\bibitem[{{Harker} {et~al.}(2009){Harker}, {Zaroubi}, {Bernardi}, {Brentjens},
  {de Bruyn}, {Ciardi}, {Jeli{\'c}}, {Koopmans}, {Labropoulos}, {Mellema},
  {Offringa}, {Pandey}, {Schaye}, {Thomas}, \& {Yatawatta}}]{Harker2009}
{Harker}, G., {et~al.} 2009, \mnras, 397, 1138

\bibitem[{{Hogan} \& {Rees}(1979)}]{Hogan1979}
{Hogan}, C.~J., \& {Rees}, M.~J. 1979, \mnras, 188, 791

\bibitem[{{H{\"o}gbom}(1974)}]{Hogbom1974}
{H{\"o}gbom}, J.~A. 1974, \aaps, 15, 417

\bibitem[{{Intema} {et~al.}(2011){Intema}, {van Weeren}, {R{\"o}ttgering}, \&
  {Lal}}]{Intema2011}
{Intema}, H.~T., {van Weeren}, R.~J., {R{\"o}ttgering}, H.~J.~A., \& {Lal},
  D.~V. 2011, \aap, 535, A38

\bibitem[{{Ishwara-Chandra} {et~al.}(2010){Ishwara-Chandra}, {Sirothia},
  {Wadadekar}, {Pal}, \& {Windhorst}}]{IshwaraChandra2010}
{Ishwara-Chandra}, C.~H., {Sirothia}, S.~K., {Wadadekar}, Y., {Pal}, S., \&
  {Windhorst}, R. 2010, \mnras, 405, 436

\bibitem[{{Jackson}(2005)}]{Jackson2005}
{Jackson}, C. 2005, \pasa, 22, 36

\bibitem[{{Jacobs}(2011)}]{Jacobs2011pc}
{Jacobs}, D.~C. 2011, private communication

\bibitem[{{Jacobs} {et~al.}(2011){Jacobs}, {Aguirre}, {Parsons}, {Pober},
  {Bradley}, {Carilli}, {Gugliucci}, {Manley}, {van der Merwe}, {Moore}, \&
  {Parashare}}]{Jacobs2011}
{Jacobs}, D.~C., {et~al.} 2011, \apjl, 734, L34

\bibitem[{{Kassim} {et~al.}(2007){Kassim}, {Lazio}, {Erickson}, {Perley},
  {Cotton}, {Greisen}, {Cohen}, {Hicks}, {Schmitt}, \& {Katz}}]{Kassim2007}
{Kassim}, N.~E., {et~al.} 2007, \apjs, 172, 686

\bibitem[{{Lane} {et~al.}(2004){Lane}, {Clarke}, {Taylor}, {Perley}, \&
  {Kassim}}]{Lane2004}
{Lane}, W.~M., {Clarke}, T.~E., {Taylor}, G.~B., {Perley}, R.~A., \& {Kassim},
  N.~E. 2004, \aj, 127, 48

\bibitem[{{Large} {et~al.}(1981){Large}, {Mills}, {Little}, {Crawford}, \&
  {Sutton}}]{Large1981}
{Large}, M.~I., {Mills}, B.~Y., {Little}, A.~G., {Crawford}, D.~F., \&
  {Sutton}, J.~M. 1981, \mnras, 194, 693

\bibitem[{{Lidz} {et~al.}(2008){Lidz}, {Zahn}, {McQuinn}, {Zaldarriaga}, \&
  {Hernquist}}]{Lidz2008}
{Lidz}, A., {Zahn}, O., {McQuinn}, M., {Zaldarriaga}, M., \& {Hernquist}, L.
  2008, \apj, 680, 962

\bibitem[{{Liu} \& {Tegmark}(2011)}]{Liu2011}
{Liu}, A., \& {Tegmark}, M. 2011, \prd, 83, 103006

\bibitem[{{Liu} {et~al.}(2009){Liu}, {Tegmark}, {Bowman}, {Hewitt}, \&
  {Zaldarriaga}}]{Liu2009}
{Liu}, A., {Tegmark}, M., {Bowman}, J., {Hewitt}, J., \& {Zaldarriaga}, M.
  2009, \mnras, 398, 401

\bibitem[{{Lonsdale} {et~al.}(2009){Lonsdale}, {Cappallo}, {Morales}, {Briggs},
  {Benkevitch}, {Bowman}, {Bunton}, {Burns}, {Corey}, {Desouza}, {Doeleman},
  {Derome}, {Deshpande}, {Gopala}, {Greenhill}, {Herne}, {Hewitt}, {Kamini},
  {Kasper}, {Kincaid}, {Kocz}, {Kowald}, {Kratzenberg}, {Kumar}, {Lynch},
  {Madhavi}, {Matejek}, {Mitchell}, {Morgan}, {Oberoi}, {Ord},
  {Pathikulangara}, {Prabu}, {Rogers}, {Roshi}, {Salah}, {Sault}, {Shankar},
  {Srivani}, {Stevens}, {Tingay}, {Vaccarella}, {Waterson}, {Wayth}, {Webster},
  {Whitney}, {Williams}, \& {Williams}}]{Lonsdale2009}
{Lonsdale}, C.~J., {et~al.} 2009, IEEE Proceedings, 97, 1497

\bibitem[{{Madau} {et~al.}(1997){Madau}, {Meiksin}, \& {Rees}}]{Madau1997}
{Madau}, P., {Meiksin}, A., \& {Rees}, M.~J. 1997, \apj, 475, 429

\bibitem[{{McQuinn} {et~al.}(2006){McQuinn}, {Zahn}, {Zaldarriaga},
  {Hernquist}, \& {Furlanetto}}]{McQuinn2006}
{McQuinn}, M., {Zahn}, O., {Zaldarriaga}, M., {Hernquist}, L., \& {Furlanetto},
  S.~R. 2006, \apj, 653, 815

\bibitem[{{Mitchell} {et~al.}(2008){Mitchell}, {Greenhill}, {Wayth}, {Sault},
  {Lonsdale}, {Cappallo}, {Morales}, \& {Ord}}]{Mitchell2008}
{Mitchell}, D.~A., {Greenhill}, L.~J., {Wayth}, R.~B., {Sault}, R.~J.,
  {Lonsdale}, C.~J., {Cappallo}, R.~J., {Morales}, M.~F., \& {Ord}, S.~M. 2008,
  IEEE Journal of Selected Topics in Signal Processing, 2, 707

\bibitem[{{Morales} \& {Hewitt}(2004)}]{Morales2004}
{Morales}, M.~F., \& {Hewitt}, J. 2004, \apj, 615, 7

\bibitem[{{Morales} \& {Wyithe}(2010)}]{Morales2010}
{Morales}, M.~F., \& {Wyithe}, J.~S.~B. 2010, \araa, 48, 127

\bibitem[{{Nayak} {et~al.}(2010){Nayak}, {Daiboo}, \& {Shankar}}]{Nayak2010}
{Nayak}, A., {Daiboo}, S., \& {Shankar}, N.~U. 2010, \mnras, 408, 1061

\bibitem[{{Oberoi} {et~al.}(2011){Oberoi}, {Matthews}, {Cairns}, {Emrich},
  {Lobzin}, {Lonsdale}, {Morgan}, {Prabu}, {Vedantham}, {Wayth}, {Williams},
  {Williams}, {White}, {Allen}, {Arcus}, {Barnes}, {Benkevitch}, {Bernardi},
  {Bowman}, {Briggs}, {Bunton}, {Burns}, {Cappallo}, {Clark}, {Corey},
  {Dawson}, {DeBoer}, {De Gans}, {deSouza}, {Derome}, {Edgar}, {Elton},
  {Goeke}, {Gopalakrishna}, {Greenhill}, {Hazelton}, {Herne}, {Hewitt},
  {Kamini}, {Kaplan}, {Kasper}, {Kennedy}, {Kincaid}, {Kocz}, {Koeing},
  {Kowald}, {Lynch}, {Madhavi}, {McWhirter}, {Mitchell}, {Morales}, {Ng},
  {Ord}, {Pathikulangara}, {Rogers}, {Roshi}, {Salah}, {Sault}, {Schinckel},
  {Udaya Shankar}, {Srivani}, {Stevens}, {Subrahmanyan}, {Thakkar}, {Tingay},
  {Tuthill}, {Vaccarella}, {Waterson}, {Webster}, \& {Whitney}}]{Oberoi2011}
{Oberoi}, D., {et~al.} 2011, \apjl, 728, L27

\bibitem[{{Ord} {et~al.}(2010){Ord}, {Mitchell}, {Wayth}, {Greenhill},
  {Bernardi}, {Gleadow}, {Edgar}, {Clark}, {Allen}, {Arcus}, {Benkevitch},
  {Bowman}, {Briggs}, {Bunton}, {Burns}, {Cappallo}, {Coles}, {Corey},
  {Desouza}, {Doeleman}, {Derome}, {Deshpande}, {Emrich}, {Goeke},
  {Gopalakrishna}, {Herne}, {Hewitt}, {Kamini}, {Kaplan}, {Kasper}, {Kincaid},
  {Kocz}, {Kowald}, {Kratzenberg}, {Kumar}, {Lonsdale}, {Lynch}, {McWhirter},
  {Madhavi}, {Matejek}, {Morales}, {Morgan}, {Oberoi}, {Pathikulangara},
  {Prabu}, {Rogers}, {Roshi}, {Salah}, {Schinkel}, {Udaya Shankar}, {Srivani},
  {Stevens}, {Tingay}, {Vaccarella}, {Waterson}, {Webster}, {Whitney},
  {Williams}, \& {Williams}}]{Ord2010}
{Ord}, S.~M., {et~al.} 2010, \pasp, 122, 1353

\bibitem[{{Paciga} {et~al.}(2011){Paciga}, {Chang}, {Gupta}, {Nityanada},
  {Odegova}, {Pen}, {Peterson}, {Roy}, \& {Sigurdson}}]{Paciga2011}
{Paciga}, G., {et~al.} 2011, \mnras, 413, 1174

\bibitem[{{Pandey}(2006)}]{Pandey2006}
{Pandey}, V.~N. 2006, PhD thesis, Raman Research Institute

\bibitem[{{Parsons} {et~al.}(2010){Parsons}, {Backer}, {Foster}, {Wright},
  {Bradley}, {Gugliucci}, {Parashare}, {Benoit}, {Aguirre}, {Jacobs},
  {Carilli}, {Herne}, {Lynch}, {Manley}, \& {Werthimer}}]{Parsons2010}
{Parsons}, A.~R., {et~al.} 2010, \aj, 139, 1468

\bibitem[{{Pedani}(2003)}]{Pedani2003}
{Pedani}, M. 2003, \na, 8, 805

\bibitem[{{Pen} {et~al.}(2009){Pen}, {Chang}, {Hirata}, {Peterson}, {Roy},
  {Gupta}, {Odegova}, \& {Sigurdson}}]{Pen2009}
{Pen}, U.-L., {Chang}, T.-C., {Hirata}, C.~M., {Peterson}, J.~B., {Roy}, J.,
  {Gupta}, Y., {Odegova}, J., \& {Sigurdson}, K. 2009, \mnras, 399, 181

\bibitem[{{Pritchard} \& {Loeb}(2011)}]{Pritchard2011}
{Pritchard}, J.~R., \& {Loeb}, A. 2011, Prog. Rep. Phys., submitted
  (arXiv:1109.6012)

\bibitem[{{Rengelink} {et~al.}(1997){Rengelink}, {Tang}, {de Bruyn}, {Miley},
  {Bremer}, {Roettgering}, \& {Bremer}}]{Rengelink1997}
{Rengelink}, R.~B., {Tang}, Y., {de Bruyn}, A.~G., {Miley}, G.~K., {Bremer},
  M.~N., {Roettgering}, H.~J.~A., \& {Bremer}, M.~A.~R. 1997, \aaps, 124, 259

\bibitem[{{Rottgering} {et~al.}(2006){Rottgering}, {Braun}, {Barthel}, {van
  Haarlem}, {Miley}, {Morganti}, {Snellen}, {Falcke}, {de Bruyn}, {Stappers},
  {Boland}, {Butcher}, {de Geus}, {Koopmans}, {Fender}, {Kuijpers},
  {Schilizzi}, {Vogt}, {Wijers}, {Wise}, {Brouw}, {Hamaker}, {Noordam},
  {Oosterloo}, {Bahren}, {Brentjens}, {Wijnholds}, {Bregman}, {van Cappellen},
  {Gunst}, {Kant}, {Reitsma}, {van der Schaaf}, \& {de Vos}}]{Rottgering2006}
{Rottgering}, H.~J.~A., {et~al.} 2006, 'Cosmology, Galaxy Formation and
  Astroparticle Physics on the Pathway to the SKA' eds. H.~R. Klockner, S.
  Rawlings, M. Jarvis, A. Taylor (ASTRON: Dwingeloo), p169

\bibitem[{{Schwab}(1984)}]{Schwab1984}
{Schwab}, F.~R. 1984, \aj, 89, 1076

\bibitem[{{Scott} \& {Rees}(1990)}]{Scott1990}
{Scott}, D., \& {Rees}, M.~J. 1990, \mnras, 247, 510

\bibitem[{{Shaver} {et~al.}(1999){Shaver}, {Windhorst}, {Madau}, \& {de
  Bruyn}}]{Shaver1999}
{Shaver}, P.~A., {Windhorst}, R.~A., {Madau}, P., \& {de Bruyn}, A.~G. 1999,
  \aap, 345, 380

\bibitem[{{Sirothia} {et~al.}(2011){Sirothia}, {Kantharia}, {Ishwara-Chandra},
  \& {Gopal-Krishna}}]{Sirothia2011}
{Sirothia}, S.~K., {Kantharia}, N.~G., {Ishwara-Chandra}, C.~H., \&
  {Gopal-Krishna}. 2011, The GMRT Sky Survey,
  {\url{http://tgss.ncra.tifr.res.in/}}

\bibitem[{{Slee}(1977)}]{Slee1977}
{Slee}, O.~B. 1977, Australian Journal of Physics Astrophysical Supplement, 43,
  1

\bibitem[{{Slee}(1995)}]{Slee1995}
---. 1995, Australian Journal of Physics, 48, 143

\bibitem[{{Swarup} {et~al.}(1991){Swarup}, {Ananthakrishnan}, {Kapahi}, {Rao},
  {Subrahmanya}, \& {Kulkarni}}]{Swarup1991}
{Swarup}, G., {Ananthakrishnan}, S., {Kapahi}, V.~K., {Rao}, A.~P.,
  {Subrahmanya}, C.~R., \& {Kulkarni}, V.~K. 1991, Current Science, Vol.~60,
  NO.2/JAN25, P.~95, 1991, 60, 95

\bibitem[{{Tasker} {et~al.}(1994){Tasker}, {Condon}, {Wright}, \&
  {Griffith}}]{Tasker1994}
{Tasker}, N.~J., {Condon}, J.~J., {Wright}, A.~E., \& {Griffith}, M.~R. 1994,
  \aj, 107, 2115

\bibitem[{{Tingay et al.}(in prep.)}]{Tingay2012}
{Tingay et al.}, S.~J. in prep.

\bibitem[{{Wilman} {et~al.}(2008){Wilman}, {Miller}, {Jarvis}, {Mauch},
  {Levrier}, {Abdalla}, {Rawlings}, {Kl{\"o}ckner}, {Obreschkow}, {Olteanu}, \&
  {Young}}]{Wilman2008}
{Wilman}, R.~J., {et~al.} 2008, \mnras, 388, 1335

\bibitem[{{Wright} {et~al.}(1994){Wright}, {Griffith}, {Burke}, \&
  {Ekers}}]{Wright1994}
{Wright}, A.~E., {Griffith}, M.~R., {Burke}, B.~F., \& {Ekers}, R.~D. 1994,
  \apjs, 91, 111

\bibitem[{{Wright} {et~al.}(1996){Wright}, {Griffith}, {Hunt}, {Troup},
  {Burke}, \& {Ekers}}]{Wright1996}
{Wright}, A.~E., {Griffith}, M.~R., {Hunt}, A.~J., {Troup}, E., {Burke}, B.~F.,
  \& {Ekers}, R.~D. 1996, \apjs, 103, 145

\bibitem[{{Zhang} {et~al.}(1997){Zhang}, {Zheng}, {Chen}, {Wang}, {Cao},
  {Peng}, \& {Nan}}]{Zhang1997}
{Zhang}, X., {Zheng}, Y., {Chen}, H., {Wang}, S., {Cao}, A., {Peng}, B., \&
  {Nan}, R. 1997, \aaps, 121, 59

\end{thebibliography}

\end{document}